\documentclass[fleqn, usenatbib, twocolumn]{aastex631}

\usepackage{newtxtext,newtxmath, textcomp}

\usepackage{csvsimple}

\usepackage[T1]{fontenc}

\usepackage{enumitem}

\usepackage{graphicx}	
\usepackage{bm}	
\usepackage{array}
\usepackage{hhline}
\usepackage{rotating}
\usepackage{longtable}
\usepackage[flushleft]{threeparttable}

\usepackage{makecell}
\usepackage{afterpage}

\begin{document}

\title[The MDW H$\alpha$ Sky Survey: Data Release 0]{The MDW H$\alpha$ Sky Survey: Data Release 0}

\author[0009-0004-7905-1755]{Noor Aftab}
\affiliation{Department of Astronomy, Columbia University \\ 550 W 120th Street \\ 
New York, NY 10027, USA}

\author[0009-0008-5025-9818]{Xunhe (Andrew) Zhang}
\affiliation{Department of Astronomy, Columbia University \\ 550 W 120th Street \\ 
New York, NY 10027, USA}

\author[0000-0002-2639-2001]{David R. Mittelman}
\affiliation{David Mittelman Observatory \\ 26 Serenity Street \\ Mayhill, NM 88339, USA}

\author[0000-0003-1235-7173]{Dennis di Cicco}
\affiliation{David Mittelman Observatory \\ 26 Serenity Street \\ Mayhill, NM 88339, USA}

\author[0000-0003-2692-2321]{Sean Walker}
\affiliation{David Mittelman Observatory \\ 26 Serenity Street \\ Mayhill, NM 88339, USA}

\author[0000-0002-6228-8244]{David H. Sliski}
\affiliation{David Mittelman Observatory \\ 26 Serenity Street \\ Mayhill, NM 88339, USA}

\author{Julia Homa}
\affiliation{Department of Astronomy, Columbia University \\ 550 W 120th Street \\ 
New York, NY 10027, USA}

\author[0009-0002-1128-2341]{Colin Holm-Hansen}
\affiliation{Department of Astronomy, Columbia University \\ 550 W 120th Street \\ 
New York, NY 10027, USA}

\author[0000-0002-1129-1873]{Mary Putman}
\affiliation{Department of Astronomy, Columbia University \\ 550 W 120th Street \\ 
New York, NY 10027, USA}

\author[0000-0003-2666-4430]{David Schiminovich}
\affiliation{Department of Astronomy, Columbia University \\ 550 W 120th Street \\ 
New York, NY 10027, USA}

\author[0009-0008-3683-197X]{Arne Henden}
\affiliation{David Mittelman Observatory \\ 26 Serenity Street \\ Mayhill, NM 88339, USA}

\author{Gary Walker}
\affiliation{David Mittelman Observatory \\ 26 Serenity Street \\ Mayhill, NM 88339, USA}

\correspondingauthor{Noor Aftab}
\email{mdw-survey@columbia.edu}

\begin{abstract}
The Mittelman-di Cicco-Walker (MDW) H$\alpha$ Sky Survey is an autonomously-operated and ongoing all-sky imaging survey in the narrowband H$\alpha$ wavelength. The survey was founded by amateur astronomers, and is presented here in its first stage of refinement for rigorous scientific use. Each field is exposed through an H$\alpha$ filter with a 3nm bandwidth for a total of four hours, with a pixel scale of {3.2} arcsec. Here, we introduce the first Data Release of the MDW H$\alpha$ Survey (Data Release 0, or DR0), spanning 238 fields in the region of Orion (\textasciitilde3100 deg${^2}$). DR0 includes: calibrated mean fields, star-removed mean fields, a point source catalog matched to Data Release 1 of the Panoramic Survey Telescope and Rapid Response System (Pan-STARRS1) and the INT Galactic Plane Survey (IGAPS), and mosaics \footnote[1]{DR0 components are available at \url{http://mdw.astro.columbia.edu}. The DR0 catalog can also be found on the AAS Journals Zenodo repository:\dataset[doi:10.5281/zenodo.12747455]{https://doi.org/10.5281/zenodo.12747455}}.  

\end{abstract}



\raggedbottom 

\section{Introduction}
\label{sec:introduction}

Since the early 19th century, visual and photographic spectroscopic observations have revealed that much of the light from nebular and diffuse regions in the Milky Way is dominated by discrete emission lines. The subsequent development of narrowband filter technologies led to monochromatic imaging that could target specific emission lines from these regions \citep{Hartmann1905}. The brightest of these lines, the Hydrogen Balmer alpha (H$\alpha$) line, is correlated with the presence of H II regions, where clouds of neutral hydrogen are ionized, largely, by ultraviolet (UV) light emitted from hot stars \citep[e.g.][]{ob_associations}. The freed electrons in these regions recombine with ionized hydrogen to emit H$\alpha$ light, which has since been the focus of wide-field imaging campaigns \citep{Courtes1951a,Courtes1951b,Sharpless1952,Morgan1955} and all-sky surveys (see Table 1). H$\alpha$ maps have revealed an extensive network of emission-line nebulae in the disk of the Milky Way, and have established the presence of ionized gas far above the plane \citep{reynolds84, dennison98, distribution_of_wim}.

The H$\alpha$ spectral line acts as a powerful tool for astronomers to study various stellar activity within the Milky Way, as well as to investigate the feedback cycle between stellar evolution and the surrounding interstellar medium. H$\alpha$ can be used to investigate the intense ionizing photons emitted from massive O stars and OB associations \citep{ob_associations, churchill02}. As stars evolve, they emit H$\alpha$ light with various stellar activity, such as flares and chromospheric activity \citep[e.g.][]{chromospheric_emission_f_g_k_dwarfs, ha_emission_of_m_dwarf_rotation, chromospheric_activity_rotation_praesepe_hyades}. When stars reach the end of their life cycle, they explode as supernovae and planetary nebulae, releasing huge amounts of energy and matter back into the interstellar medium, and often restarting the cycle \citep[e.g. see reviews by][]{ism_three_components, warm_ionized_medium}. Looking beyond the Milky Way, H$\alpha$ can even place limits on the strength of the faint extragalactic UV radiation field \citep{adams11, fumagalli17}. 

The abundant and varied structure unveiled in the H$\alpha$ wavelength makes it a popular filter choice in which to image for amateur astronomers. The MDW H$\alpha$ Survey is an all-sky imaging survey that was born from three amateur astronomers collaborating to create an image of the entire night sky in H$\alpha$ light. With a relatively fine pixel scale at 3.2 arcsec and eventual all-sky coverage, it provides ample opportunity to broaden our current knowledge of various H$\alpha$-emitting regions and sources, particularly in areas beyond the Galactic plane. 

Even in its early stages of data dissemination, the MDW Survey has proven useful in probing various objects and features. In the discovery of extensive OIII nebulosity in the sky near the M31 galaxy, the MDW Survey helped confirm there was no coincident H$\alpha$ emission \citep{m31_oiii_emission}. The survey confirmed H$\alpha$ counterparts in the discoveries of a UV arc in Ursa Major \citep{ursa_major_arc} and a new supernova remnant in Cepheus \citep{supernova_remnant_cepheus}. Additionally, mosaics from the MDW Survey have been used to investigate the Galactic halo supernova remnant G70.0–21.5, revealing unusually sharp, straight H$\alpha$ filaments \citep{galactic_halo_snr_sharp_ha_filaments}, and to investigate agreement between the Galaxy Evolution Explorer's (GALEX) far-ultraviolet measurements and MDW H$\alpha$ emission in high-latitude supernovae remnants \citep{uv_optical_emission_high_latitude_snr}. Closely following this early usage of the MDW Survey's data, \cite{julia_aas_presentation} and \cite{noor_aas_iposter} mark the first times the larger MDW Survey was presented to an academic audience - the former introducing the survey to the academic community, and the latter marking the survey's first Data Release (detailed in this paper).

The first Data Release, Data Release 0 (DR0), of the MDW H$\alpha$ Sky Survey covers 238 fields (\textasciitilde3100 ${{\rm deg}^2}$ of the sky) in the region of Orion, spanning a wide range of Galactic latitudes ($b$ ranging from $-67^\circ$ to $30^\circ$, or Declination ranging from $-30^\circ$ to $30^\circ$) which allows us to measure the survey's performance with varying H$\alpha$ nebulosity and point source crowding. This region also includes notable nebulae, such as the Orion and Rosette nebula. Section \ref{sec:data} describes the data collection for the survey, section \ref{sec:data_analysis} the data analysis for DR0, and section \ref{sec:data_products} the final data products included in this release. In the future, Data Release 1 (DR1) will span the entire Northern Hemisphere, and Data Release 2 (DR2) will cover the whole sky.

\subsection{Survey Background \& History}
The MDW H$\alpha$ Sky Survey was founded by three amateur astrophotographers: David Mittelman, Dennis di Cicco, and Sean Walker. di Cicco and Walker met as colleagues at Sky \& Telescope magazine, and subsequently worked together on H$\alpha$ imaging projects out of di Cicco's backyard in Sudbury, Massachussetts. Through mutual friends, the two were introduced to David Mittelman in 2009, a passionate amateur astronomer who encouraged the group to take these H$\alpha$ imaging projects a step further to cover the whole sky. The group began their preparations in 2014, and the MDW H$\alpha$ Sky Survey officially commenced in 2016. 

It is crucial to emphasize that the MDW Survey was originally an astrophotography endeavor, intended to map the faint wisps of H$\alpha$ nebulosity across the entire sky, deeper than had been accomplished at the time. The potential for scientific use was expected, and the founders additionally consulted astronomers at Harvard University and Dartmouth College to confirm their data acquisition methods would result in scientifically useful results. The science potential was quickly realized with several papers published using the uncalibrated early results, as mentioned earlier in the introduction.  

Significant manual effort is required in the survey: each night's data is reviewed to identify unacceptable exposures to be retaken due to inclement observing conditions. 
The founders of the survey relied heavily on commercially available software and hardware products in order to carry out their survey, which they found would be almost entirely sufficient for their goals. At the time, the commercially available observatory automation tools were not intended for projects at the MDW Survey's scale - the founders subsequently worked with authors of various observing software packages to engineer, test, and make solutions available to both the MDW Survey and the amateur community as a whole. With these substantial efforts required at the survey's outset, the idea was to refine the data collection as the survey progressed, in order to improve the survey's quality for rigorous scientific research - a process that is currently underway. See \cite{st_article} for further discussion about the survey's history.

The survey operates remotely mainly out of the David Mittelman Observatory (DMO), located in Mayhill, New Mexico, where nearly 80\% of the sky (including the Northern Hemisphere) can be mapped. DMO sits next to New Mexico Skies (NMS), a hosting site for remote telescopes that provides continued technical and logistical support for the survey. In August 2023, operations partially moved to the ObsTech Observatario El Sauce in Chile's Río Hurtado Valley to map the remainder of the Southern Hemisphere. 

\begin{figure*}
    \centering
    \includegraphics[width=1\linewidth]{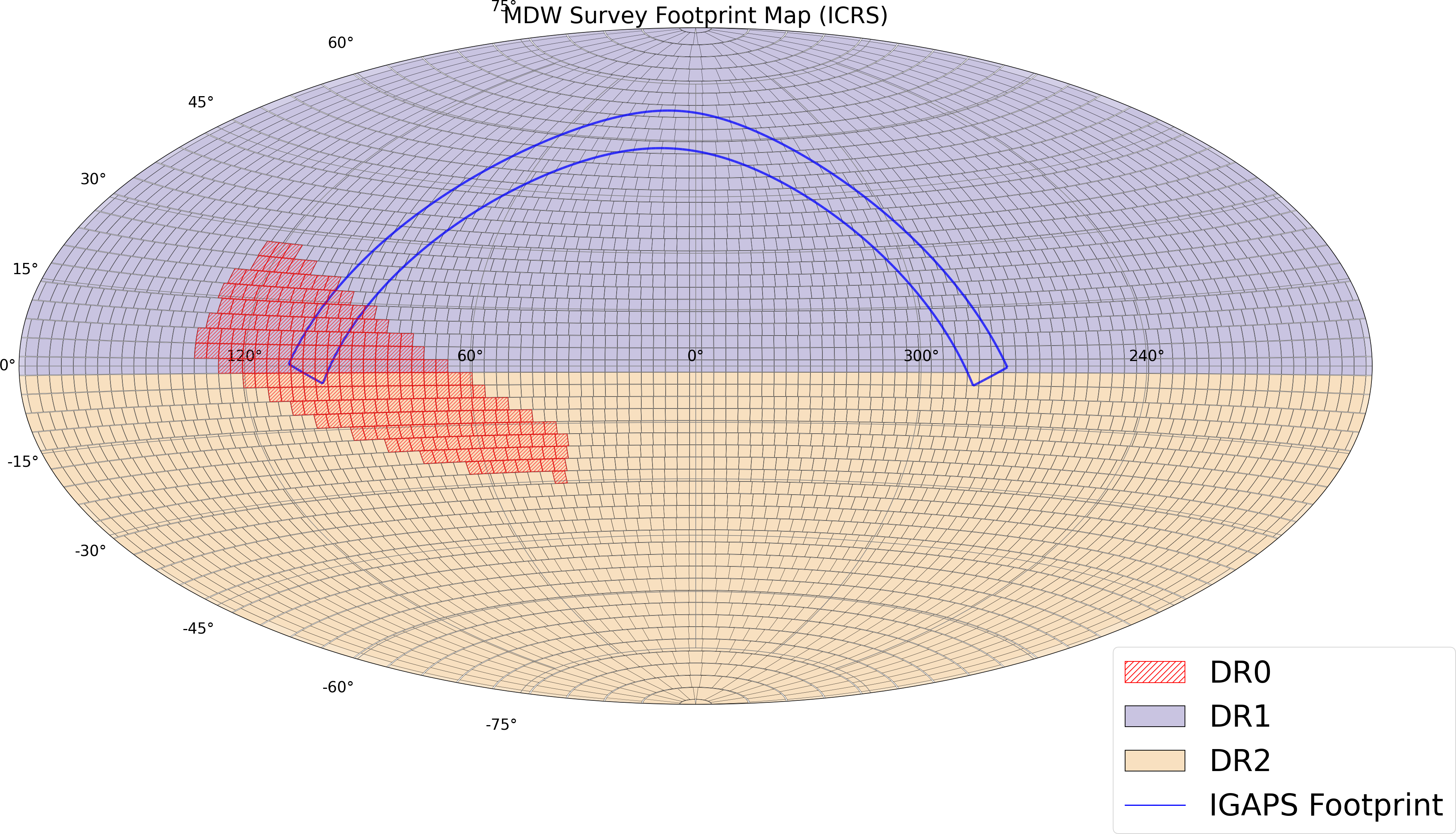}
    \caption{An all-sky map illustrating the DR0 (red), planned DR1 (purple), and planned DR2 (yellow) regions of the MDW H$\alpha$ Sky Survey. The Northern Galactic Plane covered by the IGAPS survey is shown within the blue lines. Each box represents one field, spanning $3.6 \times 3.6$ degrees. The survey will cover a total of 4120 fields, including two special fields for the North and South celestial poles. We do not include field numbers in the map due to space constraints, but the full map can be found in the online version of the journal.}
    \label{fig:all_sky_map}
\end{figure*}

\subsection{Previous Surveys}

Surveys in the H$\alpha$ wavelength (see Table~\ref{tab:survey_comparison_table}) often target specific regions. The INT Photometric H$\alpha$ Survey of the Northern Galactic Plane (IPHAS) \citep{IPHAS} and the VST Photometric H$\alpha$ Survey of the Southern Galactic Plane and Bulge (VPHAS+) \citep{VPHAS} are notable examples of H$\alpha$ surveys targeting the Galactic plane - this is an optimal location to study stellar activity and interstellar processes because of its high concentration of stars and star-forming gas.  However, mapping H$\alpha$ emission outside the Galactic plane can help us glean information from understudied star-forming regions and provide a more complete view of the Milky Way's diffuse ionized medium. The Wisconsin H$\alpha$ Mapper (WHAM) \citep{SOUTHERN_WHAM} survey provides spectroscopic information on Galactic H$\alpha$ emission across the sky;  and while the spectroscopic information provides numerous advantages, the survey is limited by its one degree spatial resolution. With its unique properties, we hope the MDW Survey will bridge this gap among H$\alpha$ sky surveys.

\section{Data}
\label{sec:data}

\begin{table*}[]
    \centering
    \begin{tabular}{|p{3cm}|p{2cm}|p{2cm}|p{2.5cm}|p{2cm}|p{2.5cm}|}
        \hline 
        \multicolumn{1}{|p{3cm}|}{} & 
        \multicolumn{1}{p{2cm}}{\textbf{SHASSA}} & 
        \multicolumn{1}{p{2cm}}{\textbf{VTSS}} & 
        \multicolumn{1}{p{2.5cm}}{\textbf{IPHAS\textsuperscript{a}/VPHAS+}} & 
        \multicolumn{1}{p{2cm}}{\textbf{WHAM\textsuperscript{b}}} & 
        \multicolumn{1}{p{2.5cm}|}{\textbf{MDW}} \\
        \hline
        \multicolumn{1}{|p{3cm}|}{\textbf{Approx. sky coverage ($\bm{{\rm deg}^2}$)} } & 
        \multicolumn{1}{p{2cm}}{20,000} & 
        \multicolumn{1}{p{2cm}}{20,000} & 
        \multicolumn{1}{p{2.5cm}}{3800 (combined)} & 
        \multicolumn{1}{p{2cm}}{All-sky} & 
        \multicolumn{1}{p{2.5cm}|}{3100 (DR0) \par All-sky (projected)} \\

        \hline
   
        \multicolumn{1}{|p{3cm}|}{\textbf{Pixel scale (arcsec/pixel)}} & 
        \multicolumn{1}{p{2cm}}{48.0} & 
        \multicolumn{1}{p{2cm}}{96.0} & 
        \multicolumn{1}{p{2.5cm}}{0.3} & 
        \multicolumn{1}{p{2cm}}{900 (0.25 deg.)} & 
        \multicolumn{1}{p{2.5cm}|}{3.2} \\
        \hline
        \multicolumn{1}{|p{3cm}|}{\textbf{FWHM of median PSF (arcsec)}} & 
        \multicolumn{1}{p{2cm}}{192.0\textsuperscript{c}} & 
        \multicolumn{1}{p{2cm}}{192.0} & 
        \multicolumn{1}{p{2.5cm}}{1.2} & 
        \multicolumn{1}{p{2cm}}{3600 (1 deg.)} & 
        \multicolumn{1}{p{2.5cm}|}{\textasciitilde6.0\textsuperscript{d}} \\
        \hline
        \multicolumn{1}{|p{3cm}|}{\textbf{Point source depth (AB mag)\textsuperscript{e}}} & 
        \multicolumn{1}{p{2cm}}{\textasciitilde15.5} & 
        \multicolumn{1}{p{2cm}}{\textasciitilde14.0} & 
        \multicolumn{1}{p{2.5cm}}{20.5} & 
        \multicolumn{1}{p{2cm}}{—} & 
        \multicolumn{1}{p{2.5cm}|}{16.0-17.0\textsuperscript{f}} \\
        
        \hline
        
        \multicolumn{1}{|p{3cm}|}{\textbf{FWHM of H$\alpha$ filter (nm)}} & 
        \multicolumn{1}{p{2cm}}{3.2} & 
        \multicolumn{1}{p{2cm}}{1.8} & 
        \multicolumn{1}{p{2.5cm}}{9.5} & 
        \multicolumn{1}{p{2cm}}{2.0} & 
        \multicolumn{1}{p{2.5cm}|}{3.0} \\
        \hline
    \end{tabular}

    \begin{tablenotes}
        \small 
        \item{$^\text{a}$ While we do not include the IGAPS survey in this table, it is worth noting that IGAPS is a merger of the IPHAS and UV-excess
Survey of the Northern Galactic Plane (UVEX) surveys.}
        \item{$^\text{b}$ Note that, per \cite{NORTHERN_WHAM}, the WHAM survey has a velocity resolution of 12 km/s with a spectral window of 200 km/s, while the rest are solely imaging surveys.}
        \item{$^\text{c}$ Per \cite{SHASSA}, this value is the effective width in SHASSA's custom-made mosaics after median filtering with a 4'x4' kernel, binning, averaging etc. }
        \item{$^\text{d}$ This is the median seeing (roughly 2 pixels) calculated over the 238 mean images in DR0, using the image segmentation packages provided by \texttt{photutils}. The median seeing varies per-field, ranging from 5.5 arcsec to 7.5 arcsec.}

        \item{$^\text{e}$ A point source depth is not readily available for SHASSA and VTSS, so we convert their reported depth in Rayleighs (0.5 R and 1 R respectively) to AB magnitudes using the following relations \citep{fesen2024deep}: {1 R (H$\alpha$) = $5.67 \times 10^{-18}$ ergs cm$^{-2}$ s$^{-1}$ arcsec$^{-2}$. m$_{AB}$ = $-2.5 \times$ log$_{10}$[$R \times 5.67 \times 10^{-18}$ ergs cm$^{-2}$ s$^{-1}$ arcsec$^{-2}$)$\times$ (median FWHM PSF)$^2$ / $\nu_{\text{bandwidth}}$ / 10$^{-23}$ ergs cm$^{-2}$ s$^{-1}$ Hz$^{-1}$] + 8.9}.}

        \item{$^\text{f}$ This reported depth is an empirical value based on the DR0 mean images, and is subject to change in future data releases as we further refine our calibration and stacking processes.}
    \end{tablenotes}
  
    \caption{A comparison of the MDW Survey with other known H$\alpha$ surveys: SHASSA \citep{SHASSA}, VTSS \citep{VTSS}, IPHAS \citep{IPHAS}, VPHAS+ \citep{VPHAS}, and WHAM \citep{NORTHERN_WHAM}. }
    \label{tab:survey_comparison_table}
\end{table*}

\subsection{Instrumentation}
\label{sec:instrumentation}

Since the survey was an expansion of ongoing H$\alpha$ imaging projects, the MDW team chose instruments largely from their prior experience in astrophotography, and the commercially available options at the time. There were three characteristics the MDW team was concerned with:

\begin{enumerate}[leftmargin=*, label=\arabic*.]
  \item \textbf{Field coverage}: A wide field of view was ideal, so as to complete the survey in a reasonable amount of time (< 10 years).
  \item \textbf{Resolution}: Since the survey images in the H$\alpha$ wavelength, a high resolution was preferred to capture the fine detail of H$\alpha$ emission. A telescope with an aperture greater than 10cm would meet this requirement.
  \item \textbf{Imaging depth}: To record the faint and subtle details in H$\alpha$, a photographic speed of f/5 or faster was ideal.
\end{enumerate}

The survey is carried out with three Astro-Physics 130GTX refractors: two for observing the Northern sky from New Mexico, and one for observing the Southern sky from Chile. The survey began with the two telescopes for the Northern Hemisphere, and both were customized such that their optical system was sealed and their observing performance as identical as possible. These modifications are more thoroughly detailed in Appendix \ref{appendix:a}. Data acquisition from the Northern Hemisphere was carried out such that each telescope recorded six exposures, resulting in an equal number of frames from each telescope. 

The telescopes are fitted with a 0.72x focal reducer resulting in an effective focal ratio of f/4.5, yielding an effective focal length of 585 mm. See Appendix \ref{appendix:a} for an in-depth comparison to other commercially available telescopes, and a more thorough discussion of the design tradeoffs considered by the team. The survey uses FLI ProLine 16803 CCD cameras, equipped with Kodak KAF-16803 sensors - this was the largest sensor available at the time for advanced amateur astronomers. Note that the KAF-16803 sensor exhibits the Residual Bulk Image (RBI) defect, where a ghost image of a prior exposure may appear in subsequent images. To mitigate this, all images are pre-flashed with LED light, filling the trap defects that cause RBI and resulting in clean science images. This process slightly increases the camera's read noise, but the image noise is typically dominated by the sky background. Due to limitations in mount tracking at the time for the required 20-minute exposures, we use Starlight Xpress Lodestar X2 cameras on separate 50 mm f/5 Borg A50 guide guide telescopes (using two second exposures). We use Astrodon 3nm narrowband H$\alpha$ filters, with a center bandwidth of 656.3 nm. We opt for this over the alternative 5nm filter because there is less unwanted emission passing through the filter, sharpening the contrast between diffuse H$\alpha$ emission and sky background.

Our telescopes in the Northern survey use the Software Bisque Paramount MX+ German equatorial mounts, and operate autonomously using the ACP Expert software by DC-3 Dreams \citep{acp_software}. For the Southern portion of the MDW Survey, we upgrade to a Planewave L350 equatorial mount that improves tracking with direct-drive motors, improves efficiency with faster slew times, and avoids doing meridian flips (unlike German equatorial mounts). 

\begin{figure}
    \centering
    \includegraphics[width=1\linewidth]{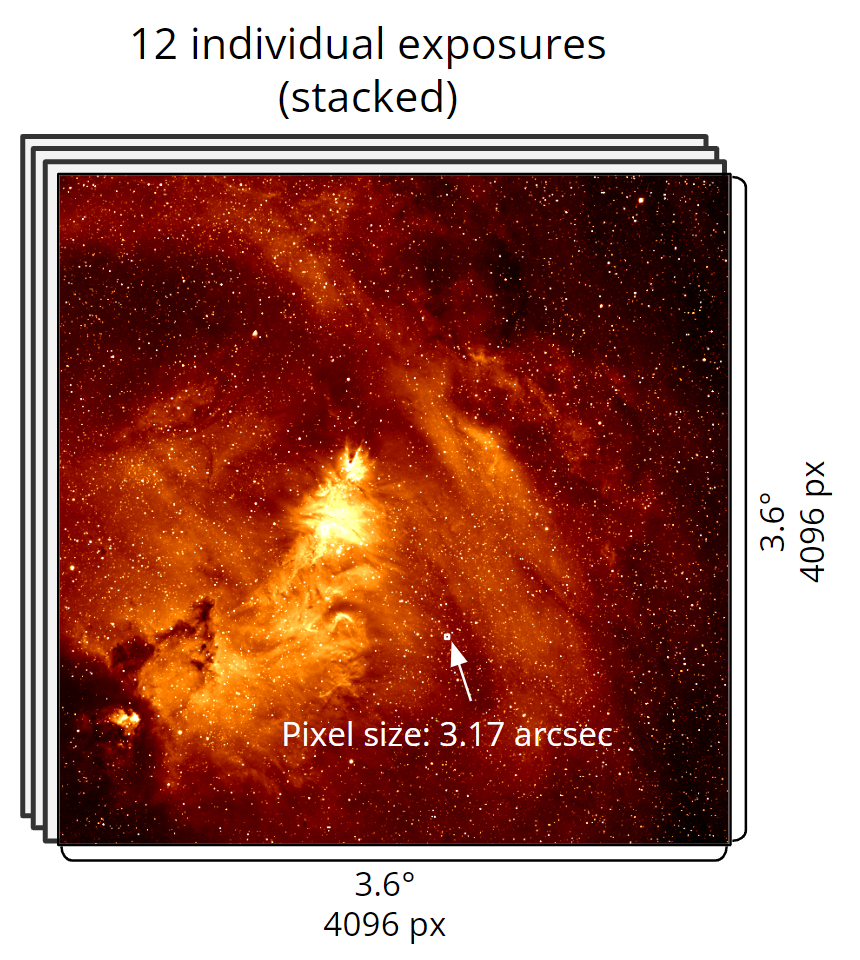}
    \caption{An illustration of the specifications of an MDW Survey field, using field 368 of DR0 as an example. The individual exposures are mean combined with sigma rejection, so as to increase the signal-to-noise ratio of the stacked field.}
    \label{fig:field_specs}
\end{figure}

\subsection{Survey Observations}
\label{sec:implementation}
\label{sec:field_specifications}

\begin{figure}
    \centering
    \includegraphics[width=1\linewidth]{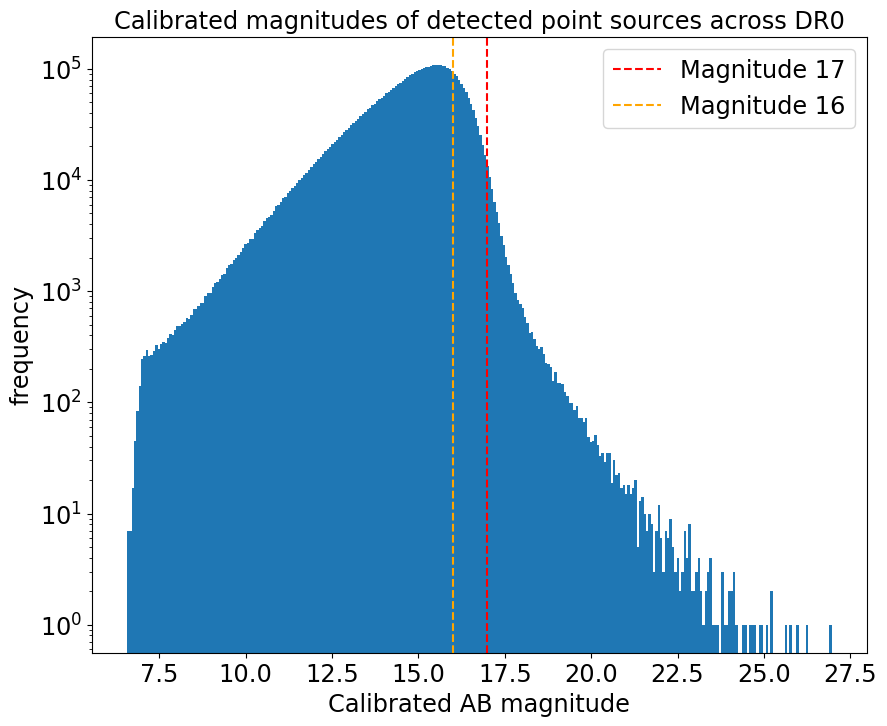}
    \caption{After calibrating our instrumental magnitudes per Section \ref{sec:photometric_calibration}, we run source detection on each DR0 image with a detection threshold of 4$\sigma$ above the image median, and apply our calculated zero-point for each field (Section \ref{sec:photometrically_calibrated_images}) to convert to calibrated AB magnitudes. Looking at where the frequency of sources drops off in this histogram, we find that the limiting magnitude for point source objects in DR0 is around 16-17 magnitudes. This value is based on the DR0 mean images, and is subject to change in future data releases, as we continue to refine our calibration and stacking processes.}
    \label{fig:mdw_point_source_depth}
\end{figure}

\begin{figure*}
    \centering
    \includegraphics[width=1\linewidth]{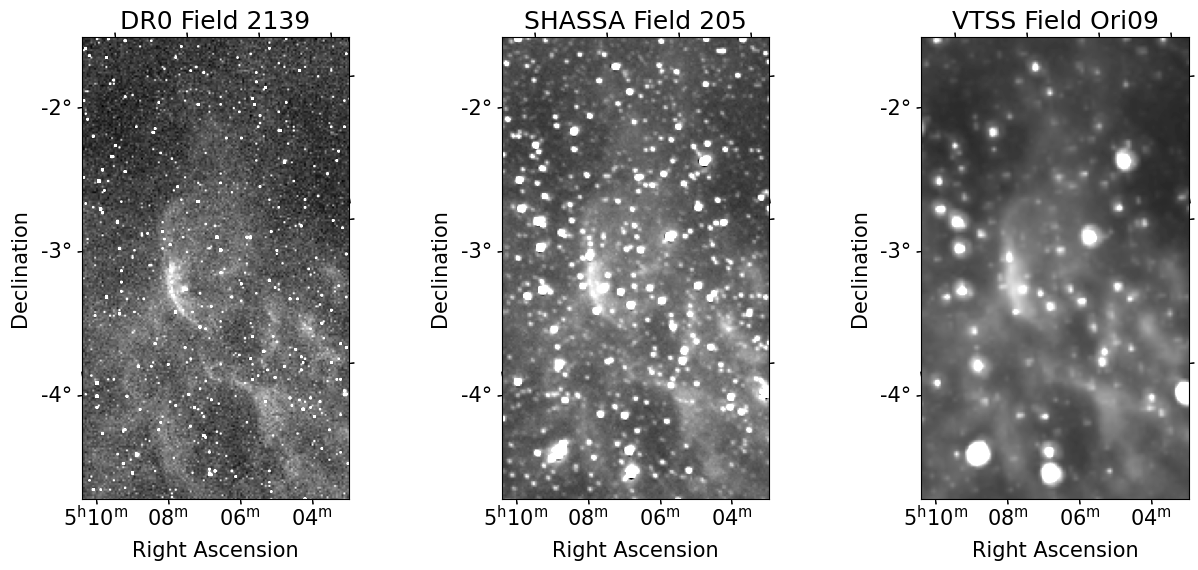}
    \caption{A plot comparing the same region of the sky as imaged in different H$\alpha$ surveys: mean-combined field 2139 of the MDW Survey (left), field 205 of SHASSA (middle), and Field Ori09 of VTSS (right). While we do not quantify the diffuse emission sensitivity of the MDW Survey, we can visually see the subtle structure revealed in the MDW field compared to the relatively low resolution SHASSA and VTSS surveys. Note that DR0 and VTSS fields have been reprojected on to SHASSA's WCS information, using \texttt{astropy}'s \texttt{reproject} package.}
    \label{fig:DR0_SHASSA_VTSS)}
\end{figure*}

Figure \ref{fig:all_sky_map} highlights DR0's footprint, and the planned DR1 and DR2 regions. Fields are numbered starting at an RA and Dec of 0$^\circ$ (field 1), with increasing RA until one "band" (a single row in the field map) is complete. This continues until the North celestial pole is reached, at which point we continue the process for the Southern Hemisphere until the South celestial pole is reached. With an overlap of \textasciitilde11 arcmin in each field, we will eventually cover the whole sky in 4120 fields, which includes two special fields each for the North and South celestial poles. Field numbers are not included in Figure \ref{fig:all_sky_map}, but the fully numbered sky map can be found in the online version of the journal. Each MDW field is mean combined with sigma rejection from 12 individual frames, each exposed for 20 minutes and resulting in a total exposure time of 4 hours. Note that the cadence of the individual frames is not uniform - some frames for the same field are taken in the same hour, while others are taken years apart. The individual frames are not included in DR0, though we intend to release them with a time series analysis from DR1 onwards.
Our combination of instruments results in a field of view that is 3.6 degrees by 3.6 degrees, with an approximate pixel scale of 3.2 arcsec/pixel (also see Figure \ref{fig:field_specs}). As shown in Figure \ref{fig:mdw_point_source_depth}, mean-combined images in DR0 have  a point source depth roughly in the 16th-17th magnitude range, although we may reach deeper magnitudes as we refine our calibration and stacking processes. While we do not quantify the diffuse emission sensitivity of our images, Figure \ref{fig:DR0_SHASSA_VTSS)} illustrates the fine detail of H$\alpha$ nebulosity in an otherwise ordinary region when imaged in the MDW Survey versus the SHASSA and VTSS surveys. Table \ref{tab:key_properties} provides a complete summary of the survey's key technical properties.

\begin{table*}
    \centering
    \begin{tabular}{>{\raggedright\arraybackslash}p{3cm} >{\raggedright\arraybackslash}p{7cm} >{\raggedright\arraybackslash}p{7cm}}
        \multicolumn{1}{l}{\textbf{Attribute}} & 
        \multicolumn{1}{p{7cm}}{\textbf{Value}} & 
        \multicolumn{1}{p{7cm}}{\textbf{Comments}} \\
        \hline
        \multicolumn{1}{l}{Telescope} & 
        \multicolumn{1}{p{7cm}}{Astro-Physics 130GTX refractor} & 
        \multicolumn{1}{p{7cm}}{0.72X Quad Telecompressor Corrector} \\
        \multicolumn{1}{l}{Camera} & 
        \multicolumn{1}{p{7cm}}{FLI ProLine 16803} & 
        \multicolumn{1}{p{7cm}}{Air is cooled to $-30^\circ$C} \\
        \multicolumn{1}{l}{Sensor} & 
        \multicolumn{1}{p{7cm}}{Kodak KAF-16803} & 
        \multicolumn{1}{p{7cm}}{1 MHz RBI Flash Readout Mode} \\
        \multicolumn{1}{l}{Filter} & 
        \multicolumn{1}{p{7cm}}{H$\alpha$} & 
        \multicolumn{1}{p{7cm}}{656.4nm (3nm bandwidth)} \\
        \multicolumn{1}{l}{No. of observing setups} & 
        \multicolumn{1}{p{7cm}}{3} & 
        \multicolumn{1}{p{7cm}}{Setups are mechanically and optically identical} \\
        \multicolumn{1}{l}{Observing location(s)} & 
        \multicolumn{1}{p{7cm}}{David Mittelman Observatory, Mayhill, New Mexico \par ObsTech SpA, Río Hurtado, Chile} & 
        \multicolumn{1}{p{7cm}}{\textasciitilde7200 ft elevation; (32.902$^\circ$ N, -105.527$^\circ$ W) \par \textasciitilde5100 ft elevation; (-30.470$^\circ$ N, -70.765$^\circ$ W)} \\
        \multicolumn{1}{l}{Exposure time} & 
        \multicolumn{1}{p{7cm}}{4 hours} & 
        \multicolumn{1}{p{7cm}}{stacked from 12 20-minute exposures} \\
        \multicolumn{1}{l}{Pixel scale} & 
        \multicolumn{1}{p{7cm}}{3.2 arcsec/pixel} & 
        \multicolumn{1}{p{7cm}}{} \\
        \multicolumn{1}{l}{Point source depth} & 
        \multicolumn{1}{p{7cm}}{16th-17th magnitude} & 
        \multicolumn{1}{p{7cm}}{DR0 mean images} \\
        \multicolumn{1}{l}{Field size} & 
        \multicolumn{1}{p{7cm}}{$3.6 \times 3.6$ degrees} & 
        \multicolumn{1}{p{7cm}}{$4096 \times 4096$ pixels} \\
        \multicolumn{1}{l}{Survey area} & 
        \multicolumn{1}{p{7cm}}{All-sky} & 
        \multicolumn{1}{p{7cm}}{DR0: \textasciitilde3100 ${{\rm deg}^2}$} \\
        \multicolumn{1}{l}{Survey footprint} & 
        \multicolumn{1}{p{7cm}}{All-sky} & 
        \multicolumn{1}{p{7cm}}{DR0: $-30^\circ \leq \text{Dec} \leq 30^\circ, 40^\circ \leq \text{RA} \leq 130^\circ$} \\
    \end{tabular}

    \caption{Key attributes of the MDW H$\alpha$ Sky Survey.}
    \label{tab:key_properties}
    
\end{table*}

\subsection{Calibration: Individual Frames}
\label{sec:calibration}

While the above sections outline information that will remain throughout the survey, the calibrations outlined in this section and the software/quality control standards detailed in Appendix \ref{appendix:b} are subject to change as we refine our data further for scientific use.

Raw individual science frames are calibrated using master dark, flat-dark, bias, and flat-field frames. We define a master frame to be a mean-combination of multiple individual frames. Roughly every two months, 10 bias and dark frames are recorded, filtered with a standard sigma rejection algorithm set at 2.5$\sigma$, and mean combined into master frames. We found 2.5$\sigma$ to be the optimal value for the calibration to remove cosmic rays and similar defects in our science frames, without sacrificing signal. 10 additional dark frames are taken matching the flat-fields' exposure, creating flat-dark frames, and are mean combined in a similar fashion. Five flat-fields are recorded each night, subtracted from the master flat-dark, and mean combined without a sigma rejection algorithm. Raw images taken with each scope are calibrated using these latest master dark, bias, and flat-field frames. After calibration, we manually inspect each frame, and reject any with strong, non-linear gradients. For fields further away from the Galactic plane, we reject frames with obviously high background levels ($\geq$ 200 counts/pixel) in non-nebulous regions. Since fields closer to the Galactic plane have high nebulosity and source density, we relax this criteria and may allow images with background levels up to 500 counts/pixel.  

Calibrated images that pass this inspection are kept, and we continue this process until each field has 12 acceptable frames. When a field is complete, we utilize the RegiStar application to register our images, and use the CCDStack2 software (\url{https://ccdware.com/ccdstack_overview/}) to refine them. We correct any mild gradients using the software's 3-point correction algorithm, which fits a plane to three background estimates and subtracts it from the image to remove the gradient. We subsequently normalize the 12 images, based on the frame with the lowest background levels and smallest average full width at half maximum (FWHM). We perform a sigma-reject algorithm on the pixels of each image, and mean combine these frames to create a mean field image.

If there are pointing issues in any of the frames, the sigma-reject algorithm may identify numerous pixels to reject. In this case, we use the MaxIm DL software (\url{https://diffractionlimited.com/product/maxim-dl/}) as an alternative, since its sigma-clip algorithm more adequately replaces the rejected pixels based on the other frames of that field. MaxIm DL additionally uses a 4-point gradient correction algorithm which proves more effective for some frames than CCDStack2's planar 3-point algorithm.

\section{Data Analysis}
\label{sec:data_analysis}


In DR0, we use and release only the mean combined images, resulting from the initial calibration described in Section \ref{sec:calibration}. For future releases, we intend to refine this initial calibration with more accurate calibrations and astrometry for individual frames, improved stacking for our mean-combined images, and additional layers of morphology and background checks. This may help increase our signal-to-noise ratio, and improve the accuracy of our source catalog.  This section describes the initial astrometry, photometry and star removal methods we applied to the DR0 mean combined images.

\subsection{Astrometry \& Point Source Extraction}
\label{sec:pse_cm}

We use the astrometry.net \citep{astrometry_net} software to plate solve our DR0 images. We first use the \texttt{photutils} \citep{photutils} Python package to detect sources in each field using image segmentation, to get an initial measure of each source's FWHM. We then extract point sources in each image using the \texttt{DAOStarFinder} class, passing in a FWHM of 3 pixels. To exclude plate-solving with anomalous pixels, we also sigma-clip the mean field images at 10$\sigma$ over 5 iterations. We then set a detection threshold of at least 3$\sigma$ of the clipped data and limit the solve to the 300 brightest sources, so the astrometry's quality is not affected by noisier sources. The source list for each image is then solved using the astrometry.net index files, with a Simple Imaging Polynomial (SIP) distortion correction of order three (also referred to as the tweak order). We additionally set a depth of 80 sources, so the astrometry.net solver checks if the 80 brightest sources can be found in a given index file before trying another. This value was found to be a reasonable balance between solving time and solution accuracy, through rough trial and error. In future releases, we intend to further explore brightness cuts and the linearity of our cameras, and the possible effects this has on our astrometric solutions.

 Following the astrometric calibration, we sought to catalog-match our point sources with other surveys in order to verify our plate solve, and to glean more data about our sources. We aimed to match our sources specifically with the Pan-STARRS1 \citep{panstarrs_surveys, PS_DR1_MAST_DOI} and IGAPS \citep{IGAPS} survey catalogs, the details of which can be found in Section \ref{sec:catalog}. We use the World Coordinate System (WCS) information in the FITS header to convert the x and y coordinates of each DR0 source into an RA and Dec coordinate. In working out the best matching technique, we first directly overlaid apertures with the Pan-STARRS1/IGAPS source coordinates falling within the DR0 images. However, this method did not do a good job because the WCS information from our initial astrometry includes distortions near the far edges of our fields. Therefore, when we use this method, the overlaid apertures would not always fully capture an MDW point source, and would partially or fully only measure background. We alternatively tried the \texttt{astropy} SkyCoord's \verb|match_to_catalog_sky| function, which finds the closest on-sky neighbor using a KDTree algorithm \citep{astropy}, and found it to be the more accurate matching method.

\begin{figure}
    \centering
    \includegraphics[width=1\linewidth]{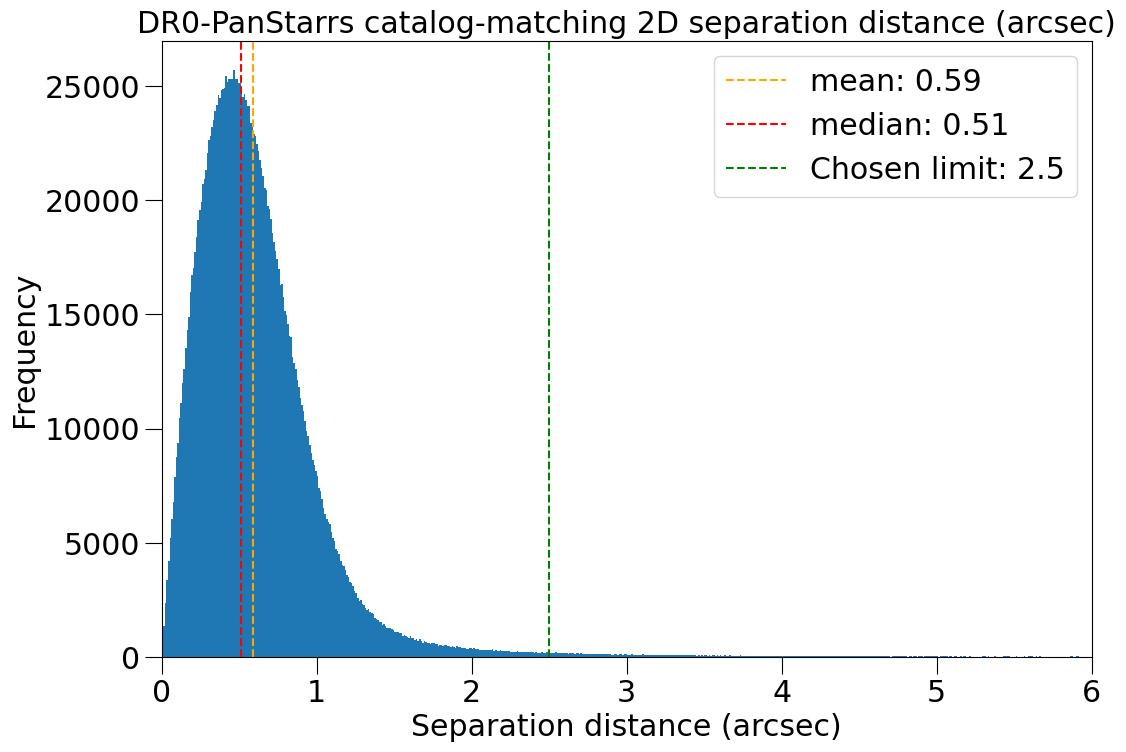}
    
    \vspace{0.3cm}
    \includegraphics[width=1\linewidth]{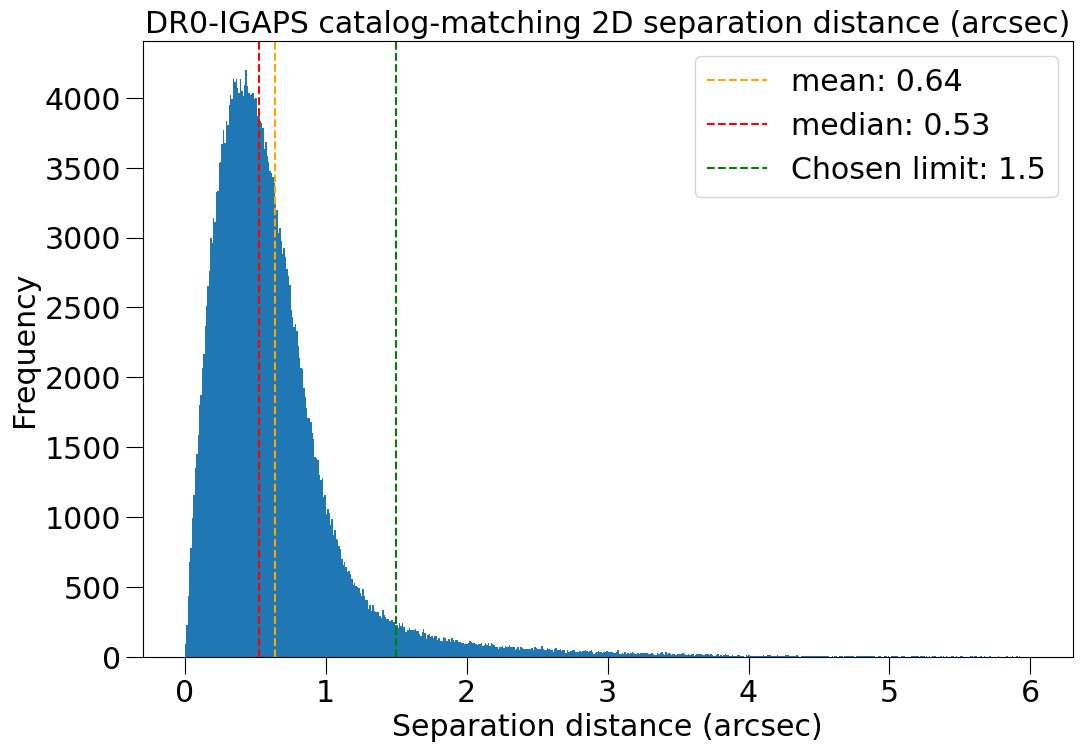}
    
    \caption{Above: Distribution of separation distances when catalog-matching DR0 sources with the Pan-STARRS1 survey. Below: Distribution of separation distances when catalog-matching DR0 sources with IGAPS. The red and yellow dotted lines are the mean and median separation distances, in arcseconds. We limit our catalog to sources with a Pan-STARRS1 match that is 2.5 arcsec or closer. IGAPS covers the Galactic plane, which may result in more mismatches because the region has high source density. Hence, we limit matching to IGAPS sources that have a separation of at most 1.5 arcsec.}
    \label{fig:ps_distance_hist}
    \label{fig:ig_distance_hist}
    \label{fig:ps_ig_distance_hist}
\end{figure}

 Figure \ref{fig:ps_ig_distance_hist} illustrates the distribution of matched MDW sources to both catalogs. We ultimately sought to limit the catalog to MDW sources with a separation of 2.5 arcsec or closer to its corresponding Pan-STARRS1 source. We felt this excludes likely mismatches while ensuring most source detections were kept. Since the Galactic plane has high source density, we are more discriminating with the IGAPS data and limit the catalog to MDW sources that are 1.5 arcsec or closer to the corresponding IGAPS source. To prevent matching with fainter (and therefore, noisier) sources, we only match to Pan-STARRS1 and IGAPS sources with a reported r magnitude brighter than 15.

\begin{figure}
    \centering
    \includegraphics[width=1\linewidth]{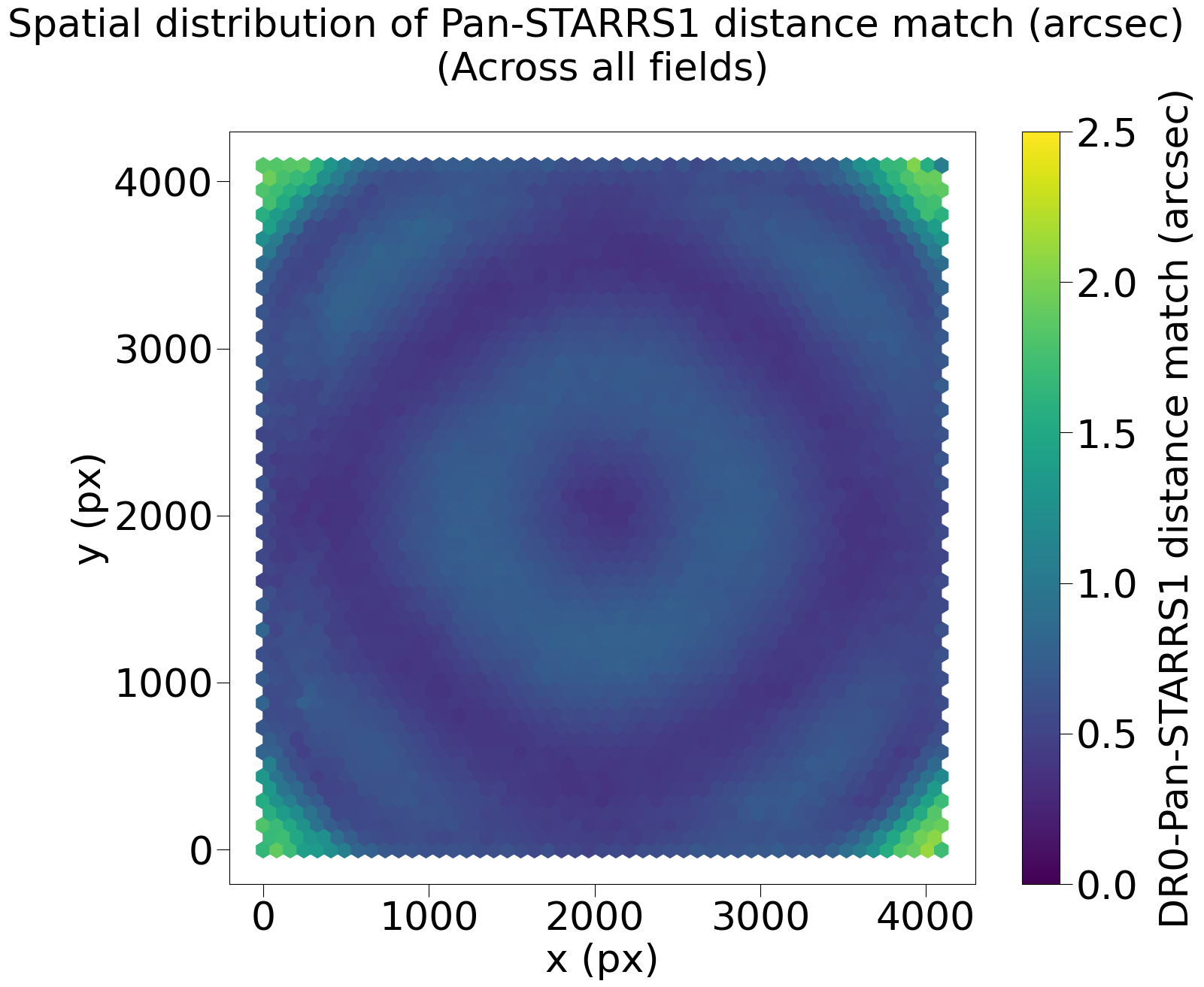}
    \includegraphics[width=1\linewidth]{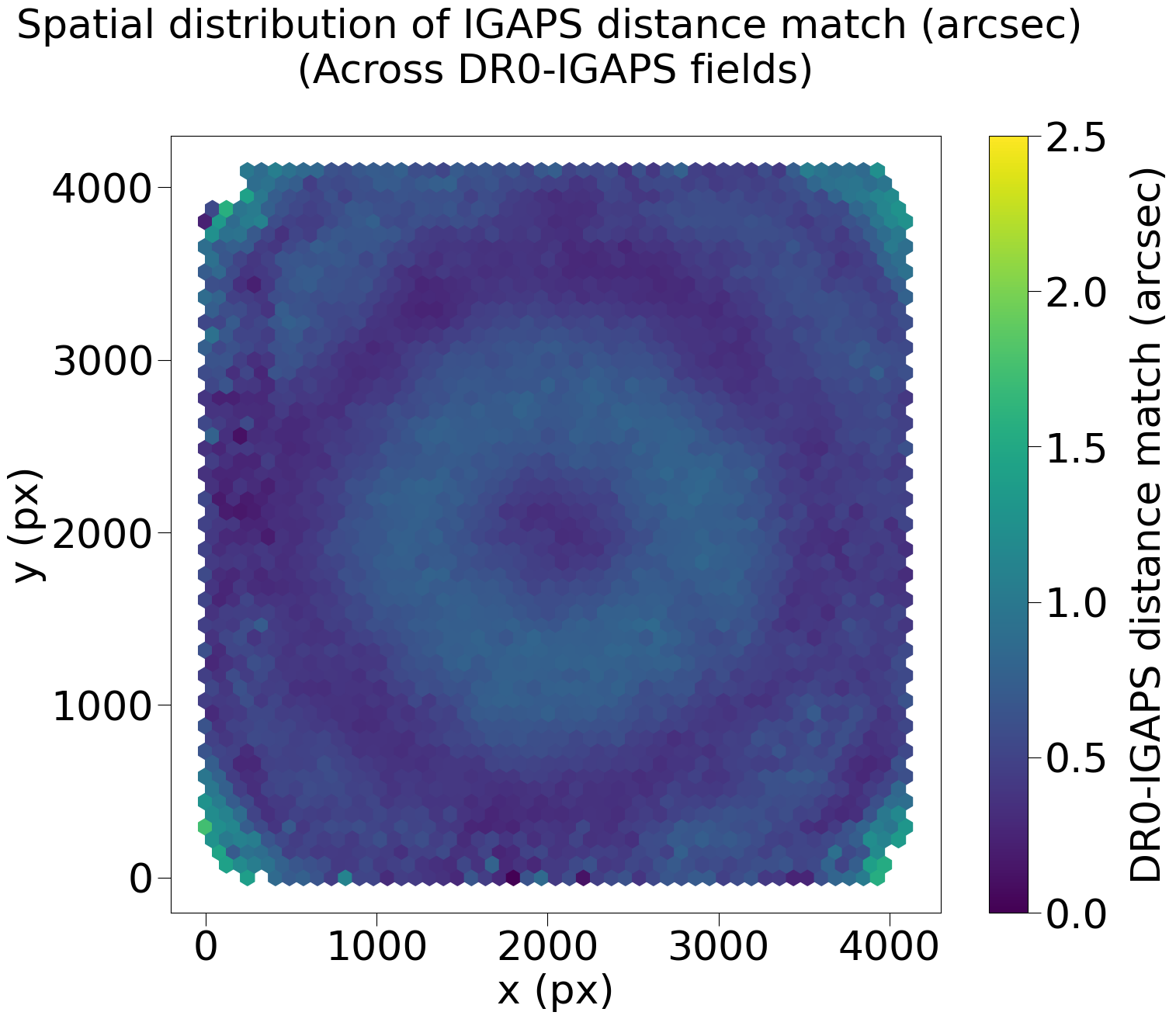}
    \caption{Above: spatial distribution of separation distances across DR0 images when catalog-matching DR0 sources with Pan-STARRS1. Below: The same distribution, using IGAPS instead. Within the bullseye pattern, we see that the separation distances are less than half a pixel, with maximum separation (< 1 pix) occurring in the field corners. To further correct these offsets, we may consider refinements to the polynomial order of our astrometric solutions in future data releases - we currently use a tweak order of three. Note that we set a separation limit of 2.5 arcsec when catalog-matching to Pan-STARRS1, and 1.5 arcsec for IGAPS.}
    \label{fig:ps_distance_2dhist}
    \label{fig:ig_distance_2dhist}
    \label{fig:ps_ig_distance_2dhist}
\end{figure}

Figure \ref{fig:ps_ig_distance_hist} shows that most matched sources have a separation of less than one arcsec which reassures us that our astrometric solutions are accurate, for the most part. However, Figure \ref{fig:ps_ig_distance_2dhist} reveals spatial patterns in the separation distances across our images. We find that sources on the higher end of match distances generally sit in the far corners of our fields, though this is usually a maximum \textasciitilde2.25 arcsec with Pan-STARRS1 sources, which is less than one pixel. Note that there are a small number of fields (< 5) with incomplete catalog coverage in the far field corners, where our detected sources did not meet the 2.5 arcsec matching limit - this likely indicates room for improvement in our astrometry. Figure \ref{fig:ps_ig_distance_2dhist} also reveals a bullseye pattern emanating from the center of our images, which may point to systematic errors. These are currently being investigated, but may be solved by iterating our astrometric solutions to use higher-order polynomial terms. Match distances within these rings are small ($\leq$ 1.25 arcsec, or less than half a pixel).

\subsection{Photometric Calibration}
\label{sec:photometric_calibration}
We perform aperture photometry on our extracted sources (Section \ref{sec:pse_cm}) using the \texttt{photutils} package. For each source, we measure the total source flux through an aperture of radius six arcsec, and the local background using an annulus with inner radius of eight arcsec and outer radius of 14 arcsec. We sigma-clip this background at 10$\sigma$ over 5 iterations, and subtract it from the source flux to calculate the net instrumental flux, $f_{\text{instrumental}}$. 
From this we can, derive the instrumental magnitude:

\begin{equation}
\label{eqn:instr_mag}
\text{MDW}_{\text{instrumental}} = -2.5\log{f_{\text{instrumental}}} 
\end{equation}

In the process of calibrating the instrumental magnitudes for sources across DR0, we calculate two offsets:
\begin{enumerate}[leftmargin=*, label=\arabic*.]
    \item Offset of MDW instrumental magnitudes to Pan-STARRS1 r magnitudes, fitted using data in the DR0 footprint. This offset is applied to the whole release. This is useful as an initial approximation.
    \item Offset of above, once-corrected magnitude, determined using the stellar locus in the $(r - H\alpha)$ vs. $(r-i)$ diagram. This value is customized to each field in DR0.
\end{enumerate}

Our first offset is used to determine the approximate range of calibrated magnitudes. It is derived by comparing the instrumental magnitudes of our DR0 sources to the r magnitude of their corresponding Pan-STARRS1 source, $\text{Pan-STARRS1}_r$. We base the offset on the r magnitude because it is the Pan-STARRS1 filter with wavelength closest to H$\alpha$. We detect 5000 sources near the center of each image, match them to Pan-STARRS1, and measure their instrumental magnitudes, per Section \ref{sec:pse_cm}. We find 5000 sources to be sufficient,  since this initial offset is only meant to get us in the ballpark of a calibrated magnitude. With these instrumental magnitudes, we model a best-fit line with an undetermined y-intercept $b$, per the equation below. Using a slope of one in this equation is sufficient, because we do not expect any non-linearities in the correspondence between true flux and the instrumental flux. Additionally, Pan-STARRS1 calibrates their instrumental magnitudes, $\text{Pan-STARRS1}_{\text{instrumental}}$, to be "top of the atmosphere", as such: they add a ballpark magnitude of \textasciitilde25, derive a zero-point magnitude from ubercal analysis, and subsequently account for flat-field corrections and airmass extinction \citep{panstarrs_calib}. These corrections are additive - not multiplicative - which further indicates that a slope of one is satisfactory.

\begin{equation}
\label{eqn:first_offset}
\text{MDW}_{\text{instrumental}} = \text{Pan-STARRS1}_r + b
\end{equation}

\begin{figure}
    \centering
    \includegraphics[width=1\linewidth]{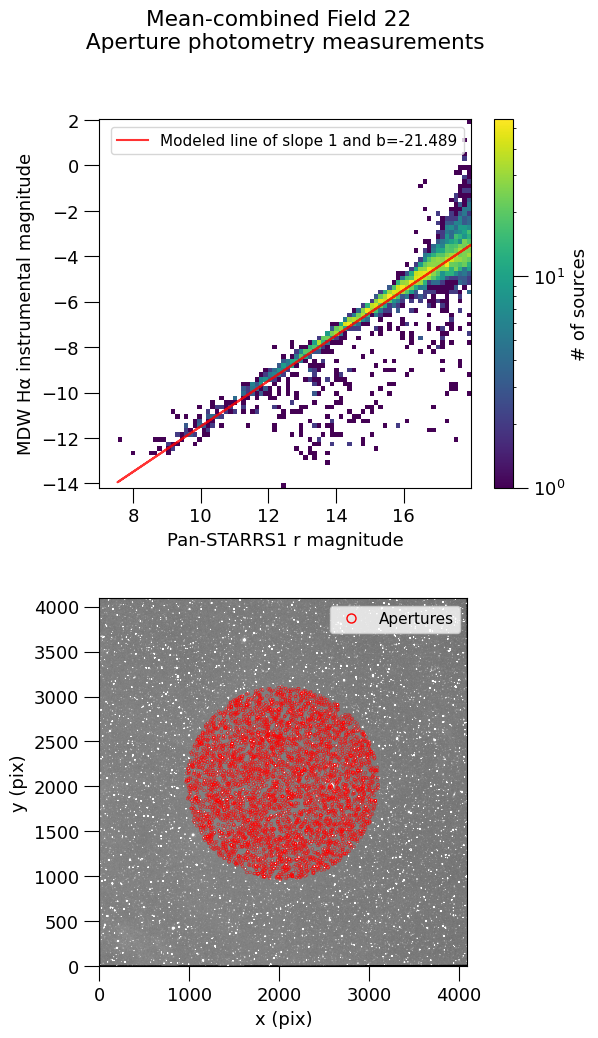}
    \caption{Top: The density plot shows our instrumental H$\alpha$ magnitudes vs. the r magnitude of the corresponding Pan-STARRS1 sources, for 5000 sources in DR0 field 22. The red line is the optimized linear model, which we then model over all DR0 fields to get the mean value of the y-intercept, b $\approx -21.5$. Bottom: We overlay 5000 apertures (red circles) at the center of field 22, from which we measure the instrumental magnitudes used for deriving the linear model in the top plot. 5000 apertures is sufficient for finding the optimal value of $b$, since it is only an initial calibration. }
    \label{fig:first_offset}
\end{figure}

We then determine a value of $b$ that minimizes the sum of the residuals across all fields, found to be approximately $-21.5$ magnitudes. See Figure \ref{fig:first_offset} for a visual representation of this process for one field. Our once-corrected H$\alpha$ magnitudes, $\text{MDW}_{\text{H$\alpha$, once\_corrected}}$ are calculated as such:

\begin{equation}
\label{eqn:once_corrected_mag}
\text{MDW}_{\text{H$\alpha$, once\_corrected}} = \text{MDW}_{\text{instrumental}} + 21.5
\end{equation}

Our second and final offset leverages color-color diagrams, specifically that of $(r-H\alpha)$ vs. $(r-i)$. These types of plots are extensively used in the IGAPS survey, and are based on the fact that stars form a locus in the color-color plane, and different types of stars sit in different areas of the plane. Various figures in \cite{IGAPS} show that IGAPS sources cluster in the region constrained by $0.5 < (\text{Pan-STARRS1}_{r} - \text{Pan-STARRS1}_i) < 0.75$, and $0 < (\text{Pan-STARRS1}_{r} - \text{Pan-STARRS1}_{\text{H}\alpha}) < 0.5$. These sources sit in between the synthetic lines provided by the IGAPS survey for the reddened and unreddened main sequence. Using this idea, we locate the locus formed by MDW sources in each field's color-color diagram, and determine roughly where we expect our (r - H$\alpha$) values to sit. We offset our once-corrected H$\alpha$ magnitude such that the (r - H$\alpha$) values for sources in this constrained (r - i) area of each field, $i$, sits at a target value of 0. This offset is described by:

\begin{eqnarray}
\label{eqn:final_mags}
\text{MDW}_{\text{H$\alpha$, final}} &=& \text{MDW}_{\text{H$\alpha$, once\_corrected}} \nonumber \\
& & + \left(\text{MDW}_{\text{H}\alpha_{\text{main\_sequence\_peak, i}}} - 0\right)
\end{eqnarray}

We sub-sample this peak before calculating the offset, such that:
\begin{enumerate}[leftmargin=*, label=\arabic*.]
    \item $|\text{Pan-STARRS1}_r - \text{MDW}_{\text{H$\alpha$, once\_corrected}}| < 5$
    \item $|\text{Pan-STARRS1}_{r} - \text{Pan-STARRS1}_i| < 2$
    \item $0 < (\text{Pan-STARRS1}_{r} - \text{Pan-STARRS1}_i) < 0.25$
\end{enumerate}

\begin{figure*}
    \centering
    \includegraphics[width=1\linewidth]{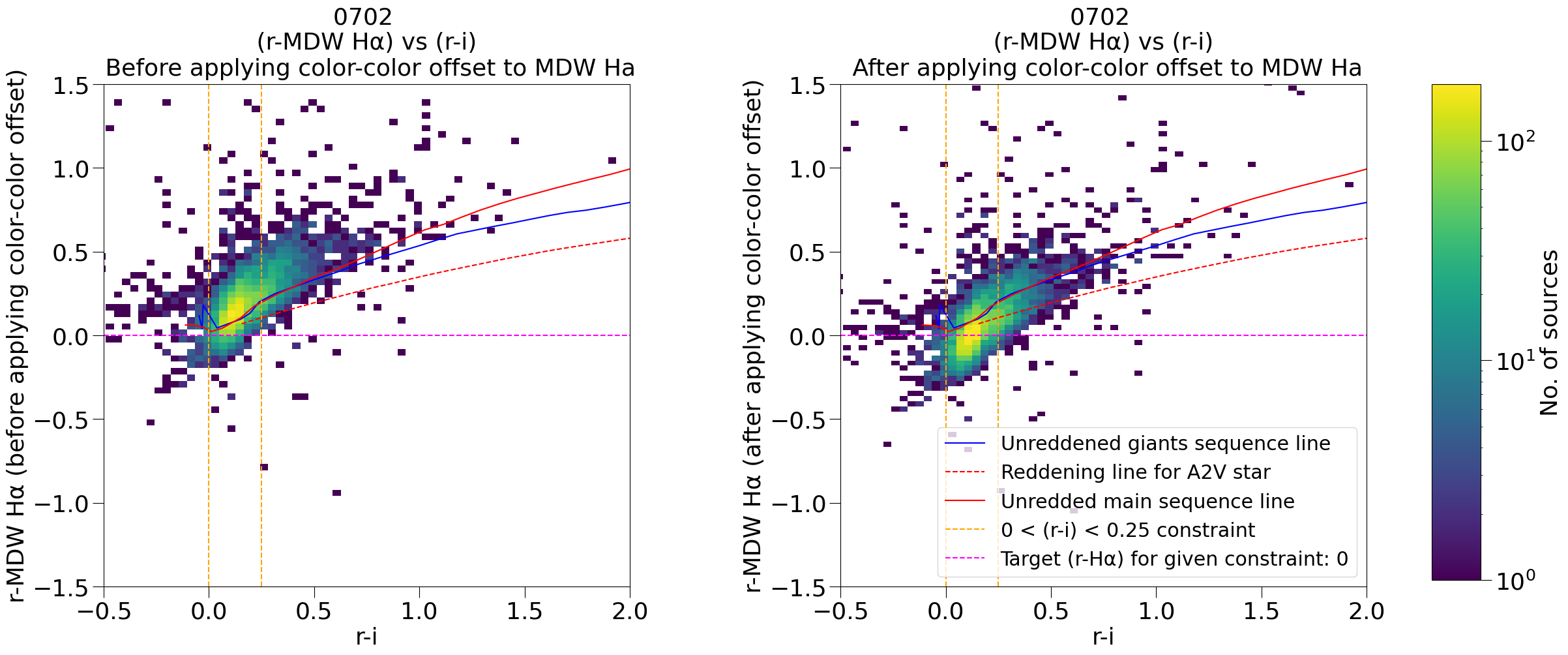}
    \caption{2D histogram of $(r-\text{MDW H}\alpha)$ vs. $(r-i)$ for DR0 field 702, before (left) and after (right) applying our second offset. Note that our stellar locus sits in the $0 < (\text{Pan-STARRS1}_{r} - \text{Pan-STARRS1}_i) < 0.25$ range, while the IGAPS stellar locus is located at $0.5 < (\text{Pan-STARRS1}_{r} - \text{Pan-STARRS1}_i) < 0.75$. This may be because IGAPS detects redder sources due to its fainter limiting magnitude (20.5 in H$\alpha$), changing the position of their locus.}
    \label{fig:color_color}
\end{figure*}

Conditions 1 and 2 ensure that the MDW H$\alpha$, Pan-STARRS1 r, and Pan-STARRS1 i magnitudes are relatively close to each other, though we add some leeway in condition 1 to account for the fact our magnitudes are not yet fully calibrated and are still off from Pan-STARRS1 r. Additionally, we believe condition 1 ensures we only use H$\alpha$ magnitudes that are (relatively) robust against photometric offsets in our peak, observed in a fraction of fields for differing reasons e.g. high background levels, varying observing conditions, systematic errors. As mentioned, the third condition constrains our peak to the $(r-i)$ range in which MDW sources form a locus.  Figure \ref{fig:color_color} shows this locus, and visualizes the application of the color-color offset. The dotted orange lines indicate where we constrain the $(r-i)$ value (condition 3) when calculating the second offset, and the dotted magenta line indicates the target $(r-\text{H}\alpha)$ value we offset towards. The solid blue, dotted red, and solid red lines are synthetic sequences used and provided by IGAPS, and assists us in choosing our target $(r-\text{H}\alpha)$ value - we see that the synthetic lines have an approximate $(r-\text{H}\alpha)$ value of 0 within our $(r-i)$ constraint, which we offset towards. Figure \ref{fig:color_color}  also reveals that our stellar locus sits in a different location of the color-color diagram than in IGAPS. Our stellar locus sits in the $0 < (\text{Pan-STARRS1}_{r} - \text{Pan-STARRS1}_i) < 0.25$ range, while the IGAP stellar locus is located at $0.5 < (\text{Pan-STARRS1}_{r} - \text{Pan-STARRS1}_i) < 0.75$ (Figure 8 in \cite{IGAPS}). The point source depth of the DR0 release is roughly 16th-17th magnitude, so our detections are likely dominated by sources that are either bluer and/or are imaged through less dust than the IGAPS survey, which has a limiting depth of \textasciitilde20.5 magnitudes in the H$\alpha$ filter \citep{IGAPS}.

\begin{figure}
    \centering
    \includegraphics[width=1\linewidth]{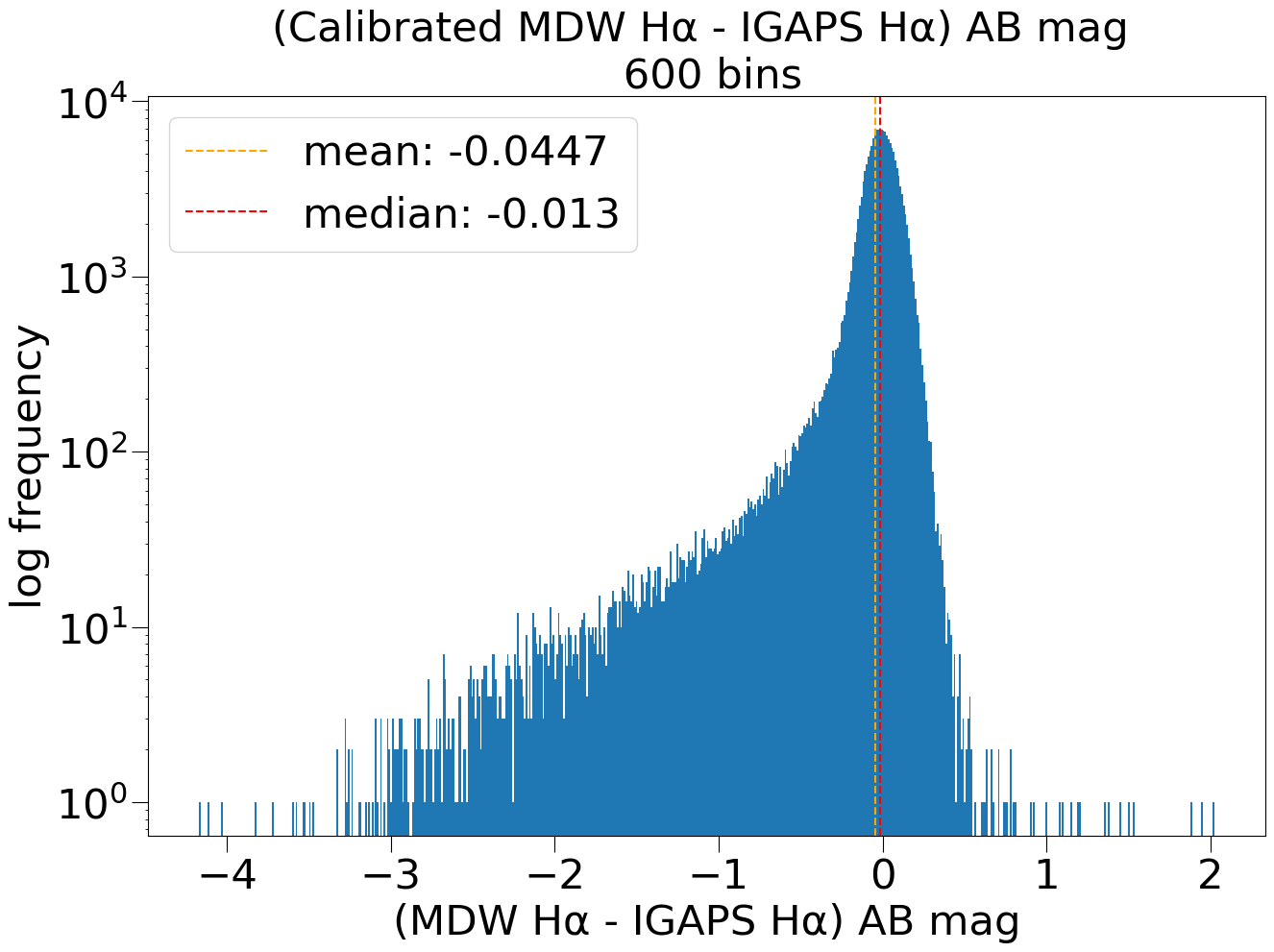}
    \caption{A histogram of the difference between our calibrated H$\alpha$ magnitudes and the corresponding IGAPS source’s H$\alpha$ AB magnitude, with a log-scale applied to the y-axis, for the region where IGAPS and DR0 overlaps. The histogram centers closely around 0, indicating that the MDW magnitudes align well with IGAPS. We do see a left-leaning tail in the histogram, indicating a number of brighter sources with fainter MDW H$\alpha$ magnitudes compared to IGAPS.}
    \label{fig:igaps_mdw_ha}
\end{figure}

\begin{figure}
    \centering
    \includegraphics[width=1\linewidth]{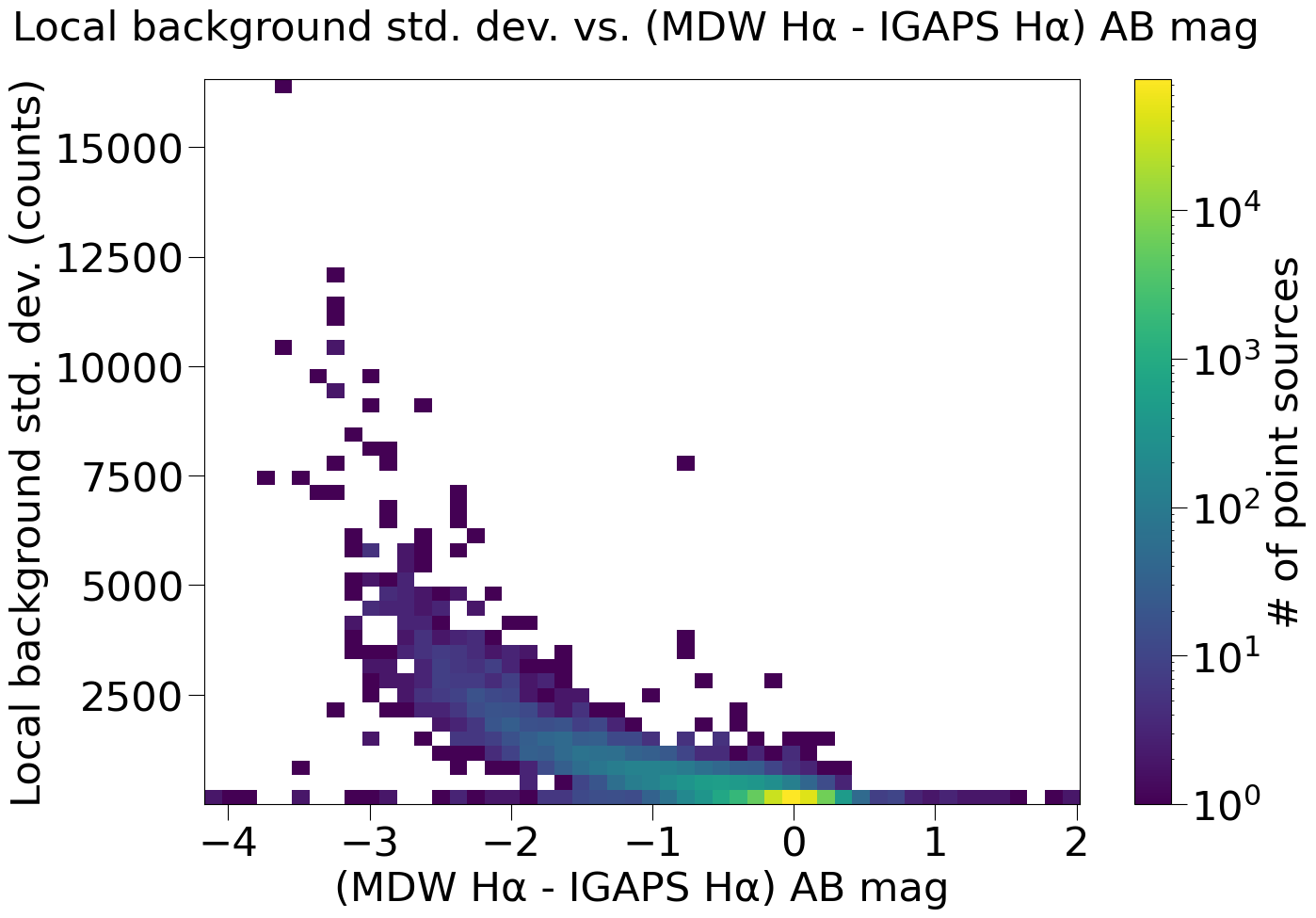}
    \caption{A 2D histogram showing the relationship between the local background standard deviation of our DR0 sources in the overlapping IGAPS region, and the difference of our calibrated H$\alpha$ magnitudes with the corresponding IGAPS H$\alpha$ magnitude.}
    \label{fig:local_bkg_std_vs_difference}
\end{figure}

\begin{figure*}[t]
    \centering
    \includegraphics[width=1\linewidth]{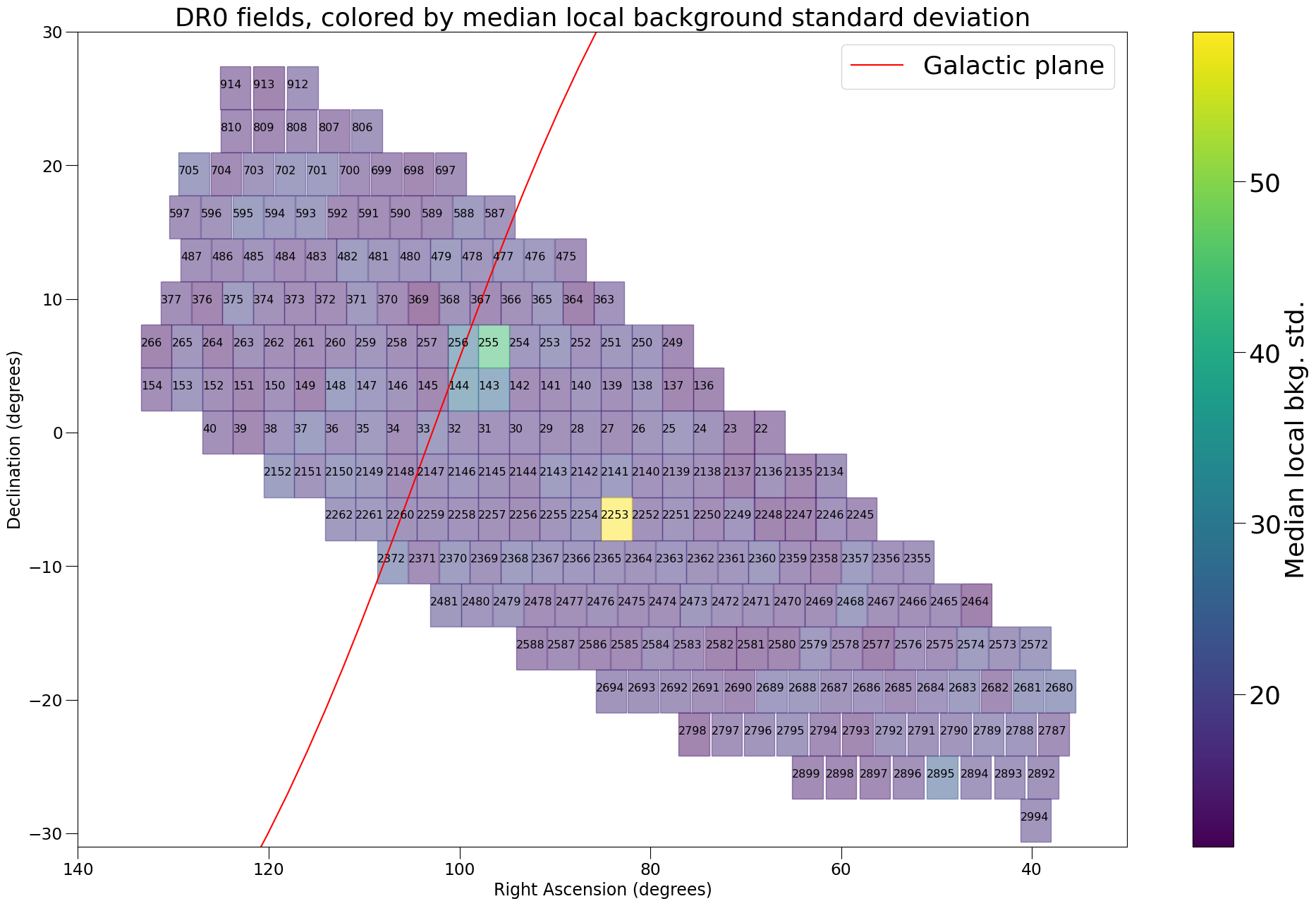}
    \caption{A plot of boxes corresponding to the width and height of each DR0 field, and numbered by field. A brighter colored box corresponds to a field with high median local background standard deviation, correlating to high nebulosity areas in the sky. The brightest field, 2253, contains the Orion nebula (M42). The Rosette nebula (NGC 2237) sits in the intersection of fields 256, 244, 144, and 143, making up the second-brightest field in the plot.}
    \label{fig:bkg_std_overlay}
\end{figure*}

We compare the calibrated H$\alpha$ magnitude of each point source to the H$\alpha$ AB magnitude of the corresponding IGAPS source, per Figure \ref{fig:igaps_mdw_ha}. We find that our values align closely with the IGAPS source catalog, with a median difference of $-0.013$ magnitudes. The distribution in Figure \ref{fig:igaps_mdw_ha} is slightly left-leaning, indicating a number of brighter sources with lower H$\alpha$ magnitude in the MDW Sky Survey than in IGAPS. 
\ref{fig:local_bkg_std_vs_difference} shows that these sources with brighter-than-expected H$\alpha$ values tend to have a high local background standard deviation. These areas of high standard deviation are subsequently pointed out in Figure \ref{fig:bkg_std_overlay}, which coincide with high nebulosity regions. For example, the Orion nebula can be found in field 2253, with the highest standard deviation among DR0 fields at \textasciitilde120 counts and plenty of expected diffuse emission in the fields around it. Similarly, the Rosette nebula sits at the intersection of fields 256, 244, 144, and 143 - all four having standard deviations well above average.

In an attempt to improve our photometric calibrations, we also compare the diffuse background of the MDW Sky Survey to IGAPS. We find a difference in intensity of roughly $-1.25  \text{ mag/arcsec}^2$, with higher background intensities (lower magnitude per solid angle) found in the MDW Survey. While the difference is measured for specific intensities, the dominant signal in these regions of higher background is presumably the H$\alpha$ emission line. We expect that this difference can be explained by correcting to line intensities, using the ratio of the H$\alpha$ filter bandwidths used by the MDW Survey and IGAPS. These filter bandwidths are 3 nm and 9.5 nm respectively (Table \ref{tab:survey_comparison_table}), which gives us a ratio of $-2.5\log_{10}(3.0 \text{ nm}/9.5 \text{ nm})\approx 1.25$. Since this ratio corresponds with the calculated difference in intensity, it indicates that our background values generally agree with the IGAPS survey. However, there are limitations in that IGAPS covers the high-emission Galactic plane, potentially negating any difference in sky noise. Our attempt at background calibration will be expanded in DR1, which has a larger overlap with the IGAPS footprint (see Figure \ref{fig:all_sky_map}). We will also explore comparing with other surveys to enhance this calibration, where possible.

\subsection{Star Removal}
\label{sec:star_removal}
\begin{figure}
    \centering
    \includegraphics[width=1\linewidth]{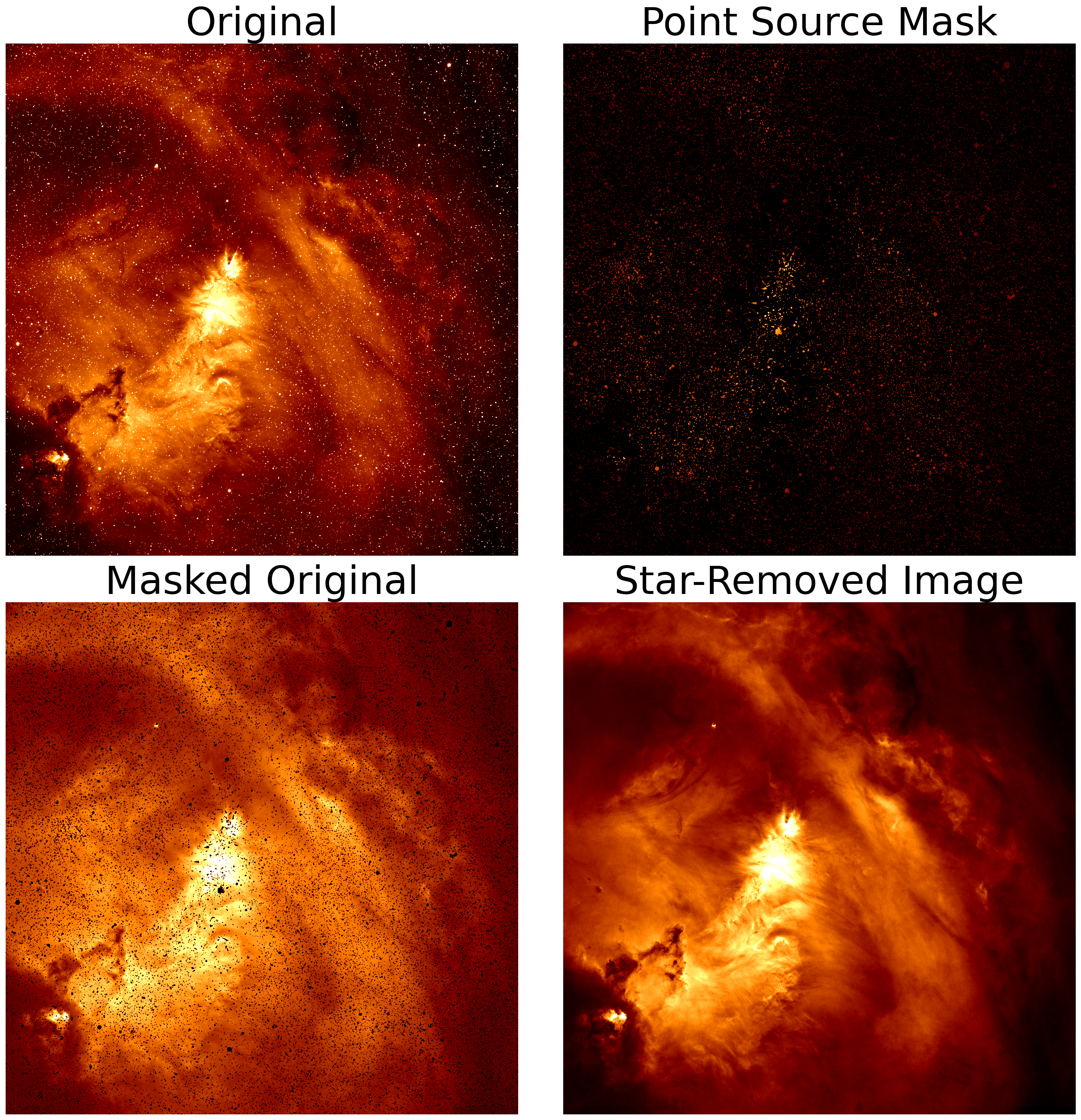}
    \caption{The above is a visual of the star-removal process for one field. The original image of field 368 is shown on the top left. We create a point source mask such that each detected source takes on the value of its local background (top right), while the detected sources are masked in the original image (bottom left). The combination of the masked image and the point-source mask yields the final star-removed image (bottom right). This process is repeated for all 238 fields in DR0.}
    \label{fig:star_removal}
\end{figure}

The faint and diffuse emission revealed by the H$\alpha$ filter can be particularly hard to identify in crowded fields. To highlight this nebulosity and make it easier to examine, we create star-masked images of each field in the DR0 region. Figure \ref{fig:star_removal} provides a visual representation of this star-removal process, which functions by masking point sources and replacing their pixels with the local background value.

To measure the value of the local background, we first create a low-resolution 2D background map through the \texttt{Background2D()} function, with a box size of 40 pixels, a filter size of 3 pixels, and using \texttt{MMMBackground()} as the 
 background estimator (both methods from the \texttt{photutils} API). The \texttt{MMMBackground()} function uses the DAOPHOT MMM algorithm \citep{Stetson1987} to calculate the local sky background in each box via mode estimation, thereby avoiding possible contamination from point sources. 
 
We then proceed in using this local background map to create a point source mask. We convolve our images to reduce detection of noisy sources, and use the \texttt{detect\_sources()} function from \texttt{photutils} to identify sources, which picks out values connected by four or more adjacent pixels with a value above a specified threshold, and returns an image with positive values at source-detected locations and zero elsewhere. In determining this mentioned threshold, we divide the fields into two distinct groups, depending on the strength of H$\alpha$ emission present. We define high emission in a field if it contains pixels at least five times the median level of its background map. For fields without high emission, we set the detection threshold to be 1.5 times the Root Mean Square (RMS) of the background map. However, we find that this threshold tends to over-mask bright nebulous regions for fields with bright emission. To compensate, we use a higher threshold of 3.5 times the RMS of the background map instead. This higher threshold leaves out extremely faint stars from the final mask but preserves the nebulosity, which is the main priority for this release component.

This point source mask is first converted to a binary mask, with a value of one at source-detected locations and zero elsewhere. We slightly increase the size of the largest masks, to help mitigate rings that were found to otherwise form around larger stars, due to them being slightly under-masked by \texttt{detect\_sources()}. This mask is then multiplied by the background map so that the source masks take the value of their local background. Finally, we take the original image and set to zero every pixel that contains a mask in the masked image, effectively creating "holes" at each source location. This modified image and the masked image are added together to create the star-masked image.

\section{Data Products}
\label{sec:data_products}

All DR0 data products (point source catalog, mean combined images, and mosaics) can be accessed on our website, \url{http://mdw.astro.columbia.edu}, including additional resources like field previews, a search function for specific coordinates of interest, quality analyses of our catalog, and contact information for the MDW team. We plan to release the 12 individual frames of each field from DR1 onwards.

\subsection{Catalog}
\label{sec:catalog}
DR0 includes a catalog of \textasciitilde1.94 million point sources matched with the Pan-STARRS1 survey, of which a further subset of \textasciitilde${160,000}$ sources  are matched with IGAPS. See Section \ref{sec:pse_cm} for full details on the point source extraction. For each source, the catalog contains information on its coordinates (RA and Dec), calibrated magnitude, instrumental flux and magnitude, non-H$\alpha$ filter magnitudes from the Pan-STARRS1/IGAPS survey catalogs, and local background information in counts. See Table \ref{tab:catalog_columns} for a description of each column in the catalog.

\begin{figure*}[t]
    \centering
    \includegraphics[width=0.8\linewidth]{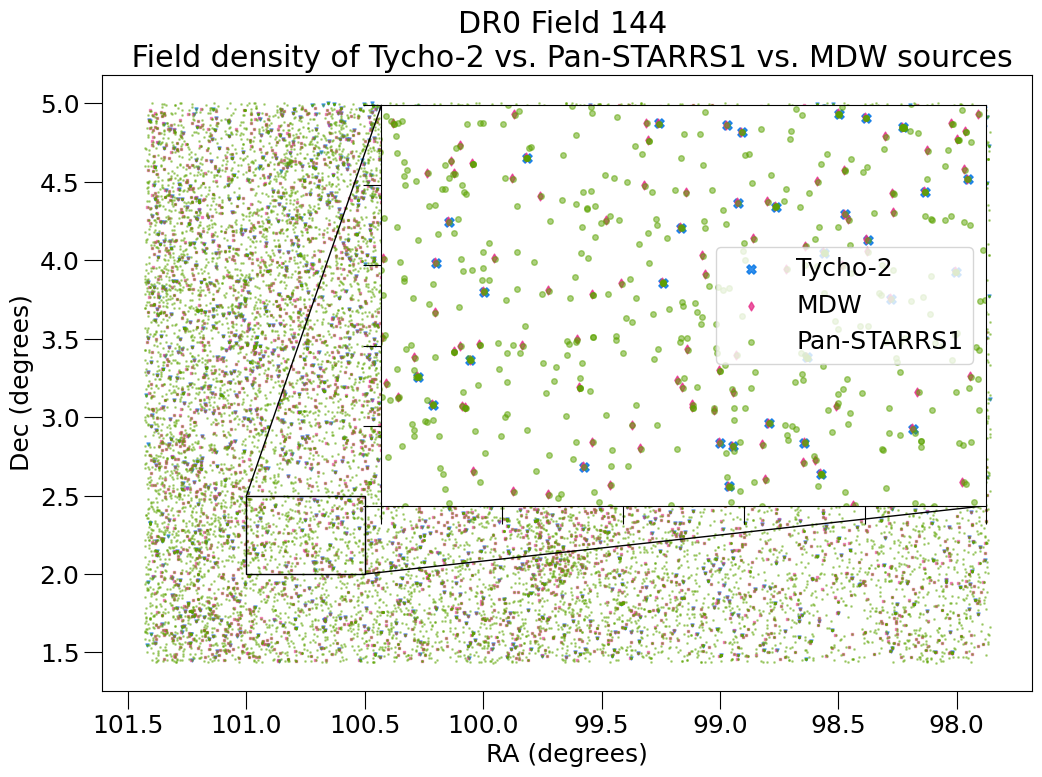}
    \caption{A comparison of field coverage and density for the Tycho-2 catalog, Pan-STARRS1 catalog, and MDW source detections, for DR0 field 144. Note that field 144 sits in the Galactic plane, where source density is particularly high. We see that the Tycho-2 catalog is relatively sparse, indicating Pan-STARRS1 as the preferred choice when creating the DR0 catalog. We also see that the MDW field has a field density that sits in the middle of Tycho and Pan-STARRS1.}
    \label{fig:tycho_vs_ps}
\end{figure*}

We initially considered using the Tycho-2 catalog \citep{tycho} of the 2.5 million brightest stars but found, upon querying with Vizier, that it is not dense enough to create an extensive DR0 source catalog, compared to Pan-STARRS1 (see Figure \ref{fig:tycho_vs_ps}). In the future, we may also explore matching our source catalog to Gaia's Data Release 3 \citep{gaia_mission, gaia_dr3}.

We also catalog-match our sources to the IGAPS survey of the Galactic plane, for the IGAPS region overlapping with DR0. This is a high resolution survey with a pixel scale of 0.33 arcsec/pixel, and has source information in broadband g, r, i filters (calibrated to Pan-STARRS1) and the narrowband H$\alpha$ and UV filters. This provides a standard to evaluate the correctness of our calibrated H$\alpha$ magnitude in the overlapping region. Figure \ref{fig:igaps_mdw_ha} showcases this comparison, revealing a median difference of $-0.013$ magnitudes between our calibrated H$\alpha$ magnitudes and the H$\alpha$ AB magnitude of the matched IGAPS source.

\begin{figure}
    \centering
    \includegraphics[width=1\linewidth]{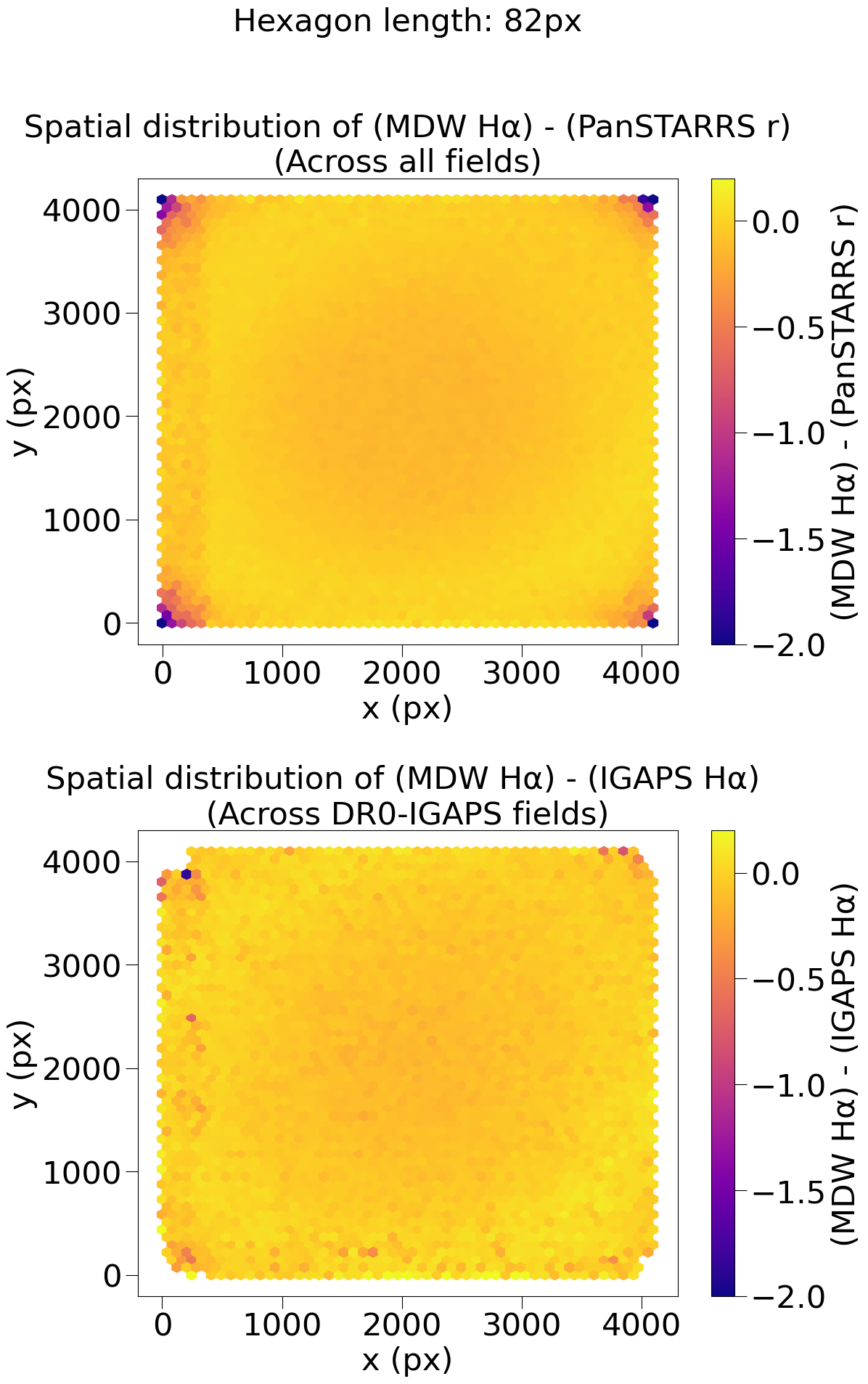}
    \caption{Above: A 2D histogram for ($\text{MDW}_{\text{H$\alpha$}} - \text{Pan-STARRS1}_{\text{r}}$), using the entire DR0 catalog.  Below: A similar histogram of all ($\text{MDW}_{\text{H$\alpha$}} - \text{IGAPS}_{\text{H$\alpha$}}$) values, for the subset of the DR0 catalog with matched IGAPS sources. In both figures, we see DR0 sources with slightly lower magnitudes near the field middles (a difference of \textasciitilde-0.25). We also notice a tendency for our DR0 sources in the field corners to have lower magnitudes than Pan-STARRS1 (differences up to \textasciitilde-2). We suspect these offsets are related to the patterns of larger separation distances in catalog-matching, which could imply mismatches, as discussed in Figure \ref{fig:ps_ig_distance_2dhist}. }
    \label{fig:all_fields_2d_hist_colors}
\end{figure}

\begin{figure}
    \centering
    \includegraphics[width=1\linewidth]{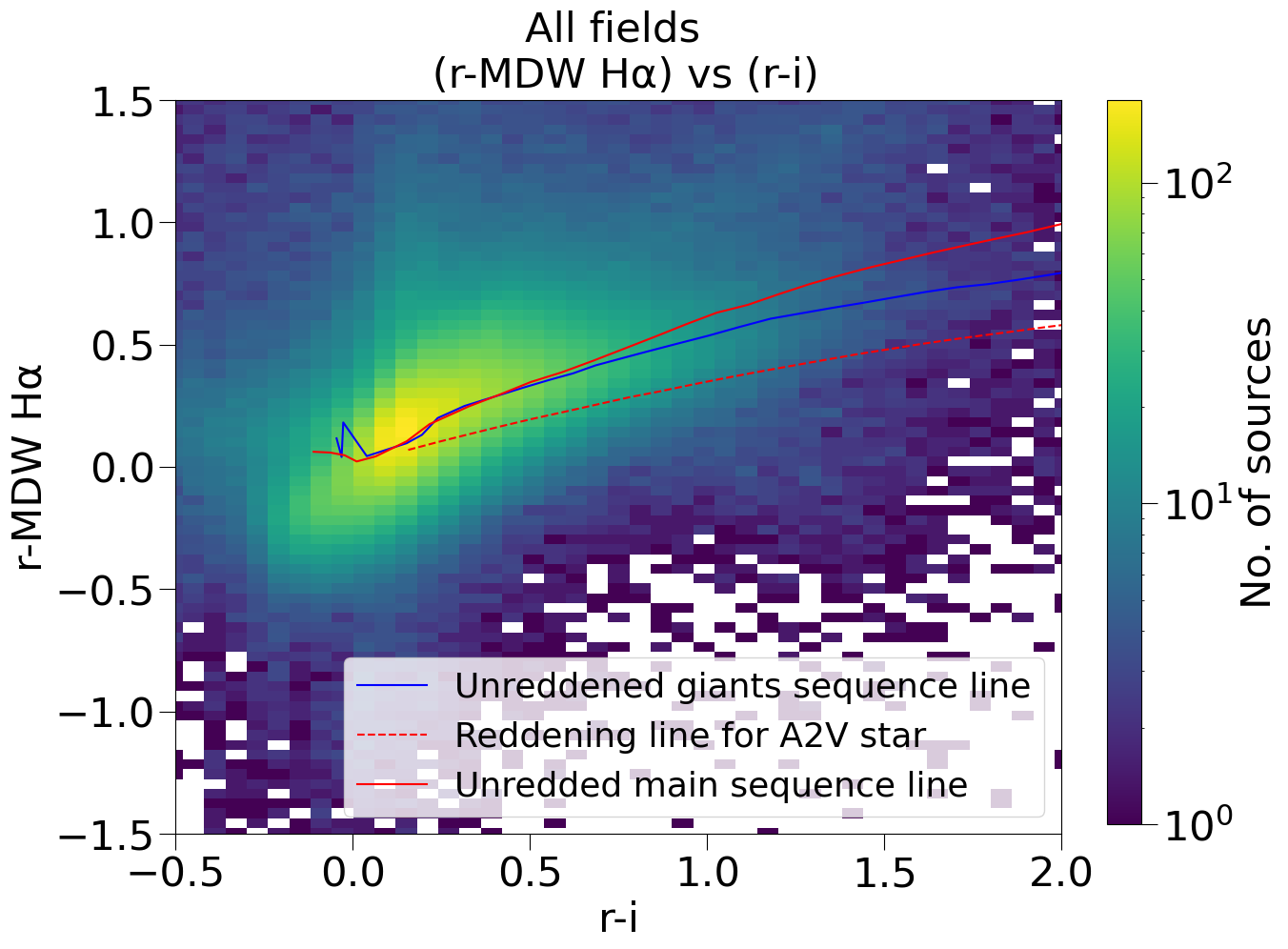}
    \caption{Pictured above is our color-color diagram of ($\text{Pan-STARRS1}_{\text{r}} - \text{MDW}_{\text{H$\alpha$}}$) vs. ($\text{Pan-STARRS1}_{\text{r}} - \text{Pan-STARRS1}_{\text{i}}$). We observe a stellar locus in the plane, with an extended feature to the right that roughly aligns with the synthetic giant sequence (solid blue line) provided by the IGAPS survey. The solid and dotted red lines are also synthetic sequences used and provided by IGAPS, which assists us in visualizing our calibration - we calibrate to the IGAPS stellar locus, which sits between the the solid blue/red and dotted red lines (Figure 8 in \cite{IGAPS}).}
    \label{fig:all_fields_color_color}
\end{figure}

We provide supplementary quality analysis (QA) figures on the MDW website (\url{http://mdw.astro.columbia.edu}) , fleshing out the catalog's overall and per-field performance for DR0. These supplemental figures include a set of 1D histograms showing the distribution of ($\text{MDW}_{\text{H$\alpha$}} - \text{IGAPS}_{\text{H$\alpha$}}$) values. Figure \ref{fig:igaps_mdw_ha} shows the overall distribution of this difference for all DR0 fields overlapping with IGAPS, whose footprint is limited to the Galactic plane. We find that the MDW calibrated H$\alpha$ magnitudes corroborate closely with the matched IGAPS sources, with a median difference of $-0.013$ magnitudes. We also include 2D histograms of the separation distance (in pixels) found during catalog-matching between DR0 source detections and the Pan-STARRS1 catalog. Figure \ref{fig:ps_distance_2dhist} reveals a bullseye distribution across DR0 fields, with the largest separation distances tending to occur in the field corners. We suspect this pattern may be related to the higher-order terms used in our astrometry, which is something we intend to investigate further. We include 2D histograms showing the spatial distribution across fields of the ($\text{MDW}_{\text{H$\alpha$}} - \text{IGAPS}_{\text{H$\alpha$}}$) and ($\text{MDW}_{\text{H$\alpha$}} - \text{Pan-STARRS1}_{\text{r}}$) colors. Figure \ref{fig:all_fields_2d_hist_colors} illustrates this for DR0 overall, with an apparent bias toward brighter MDW sources in the field corners, and a weaker  pattern of brighter MDW sources near the center. This may be related to the accuracy of our astrometric solution (the bullseye pattern and in field corners discussed earlier) possibly causing mismatches between DR0 and Pan-STARRS1/IGAPS sources. Additionally, we include color-color diagrams of ($\text{Pan-STARRS1}_{\text{r}} - \text{MDW}_{\text{H$\alpha$}}$) vs. ($\text{Pan-STARRS1}_{\text{r}} - \text{Pan-STARRS1}_{\text{i}}$).  Figure \ref{fig:all_fields_color_color} is an example illustrating this color-color plane for the entire DR0 regions, using our calibrated H$\alpha$ magnitudes. We see that our point sources form a stellar locus around r-MDW H$\alpha$ $\approx$ 0, with an extended tail of redder sources roughly aligning with the synthetic giant sequence provided by the IGAPS survey. Both of these are expected, though our data is spread out far more than the plots in \cite{IGAPS}, indicating room for improvement in DR1. Finally, we include a set of scatterplots describing the DR0 catalog's field coverage. As mentioned in Section \ref{sec:pse_cm}, a small number of fields (< 5) have sparse catalog coverage in the field corners. This is likely because of deficiencies in the DR0 astrometry, preventing us from catalog-matching to sources within a reasonable separation distance for some fields. 

\subsection{Images}
\label{sec:images}

DR0 includes 238 fields, each spanning $3.6^\circ \times 3.6^\circ$ and covering a total of \textasciitilde3100 ${{\rm deg}^2}$ of the sky. All fields in DR0 are astrometrically calibrated, with WCS information included in their headers, and mean-combined from 12 individual frames (to be released in future data sets). For each field, we include 3 types of images: \\

\noindent\textbf{{Instrumental Images}}
\label{sec:instrumental_images}

DR0 includes a set of instrumental images, in units of counts. To assist users in calibrating measured instrumental magnitudes, we include the \verb|MAGZP| (magnitude zero-point) keyword in each fields' FITS header. This is calculated from the per-field zero-point magnitude offset described in Section \ref{sec:photometric_calibration}, and can be used to derive calibrated H$\alpha$ magnitudes per the following equation: 

\begin{equation}
\label{eqn:magzp}
\text{mag(MDW)} = -2.5\log{f_{\text{instrumental}} [\text{counts}]} + {\verb|MAGZP|}_i
\end{equation}

\noindent\textbf{{Photometrically Calibrated Images}}
\label{sec:photometrically_calibrated_images}

DR0 includes photometrically calibrated images, in units of Janskys (Jy). In the process of image calibration, we first find the zero-point (ZP) instrumental counts corresponding to the zero-point magnitude. This can be found by determining what counts give us a calibrated H$\alpha$ magnitude of 0 in Equation \ref{eqn:magzp}, resulting in the following: 

\begin{equation}
\label{eqn:counts_zp}
{\texttt{MAGZP}_i = 2.5\log{f_{\text{ instrumental, ZP, i}}}}
\end{equation}

We know the zero-point magnitude must also follow the AB magnitude formula:

\begin{equation}
\label{eqn:flux_zp}
{\texttt{MAGZP}_i = -2.5\log{\frac{f_{\text{ physical, ZP, i}}}{3631 \text{ [Jy]}}}}
\end{equation}

By equating equations \ref{eqn:counts_zp} and \ref{eqn:flux_zp}, we can calculate the zero-point physical flux for a given field, $f_{\text{ physical, ZP, i}}$, as such:

\begin{equation}
\label{eqn:physical_flux_zp}
{f_{\text{ physical, ZP, i}} = \frac{3631 \text{ [Jy]}}{f_{\text{ instrumental, ZP, i}}}}
\end{equation}

We subsequently multiply our instrumental FITS data with this physical flux zero-point, thereby converting each field image from units of uncalibrated, instrumental counts to calibrated Jy. In this way, the flux measured when performing aperture photometry, $f_{\text{measured}}$, directly results in a calibrated physical flux that can be used to calculate the  H$\alpha$ AB magnitude:

\begin{equation}
\label{eqn:calib_mag}
\text{mag(MDW)} = -2.5\log{\frac{f_{\text{measured} } \text{ [Jy]}}{3631 \text{ [Jy]}}}
\end{equation}

\noindent\textbf{{Photometrically Calibrated Star-Removed Images}}
\label{sec:star_removed_images}

We include a set of point source-removed (or star-removed) images to highlight the nebulosity and diffuse emission revealed by the H$\alpha$ filter. See Section \ref{sec:star_removal} for a technical explanation of the star-removal process. These images are photometrically calibrated and in units of Jy, created and used in the same way as the calibrated star-included images.

\subsection{Mosaics}
\label{sec:mosaics}

We create mosaics of multiple fields to provide uniform and broad views of extended H$\alpha$ emission, providing an opportunity to study large-scale structures at high resolution. We intend to release larger mosaics as we release new data sets, culminating in a full-sky mosaic of H$\alpha$ emission over all Galactic latitudes in DR2. 

We use MontagePy \citep{montagepy} as the foundation of our mosaicking workflow. For each mosaic, we determine the center and size of the desired mosaic, and collect calibrated data for all fields  within these dimensions. We then determine a target frame template based on this size. For mosaics of smaller regions, the chosen template does not make a significant difference, so we often use the default gnomonic tangent plane projection. For mosaics that are larger, or at higher Declinations, we must be more cautious because different projections can distort the borders of a mosaic drastically. In these scenarios, we often use Hammer-Aitoff and Cartesian projections. We then reproject each field using the chosen projection. In certain cases we also increase the pixel scale to reduce output file sizes and improve run speed.

\begin{figure}
    \centering
    \includegraphics[width=1\linewidth]{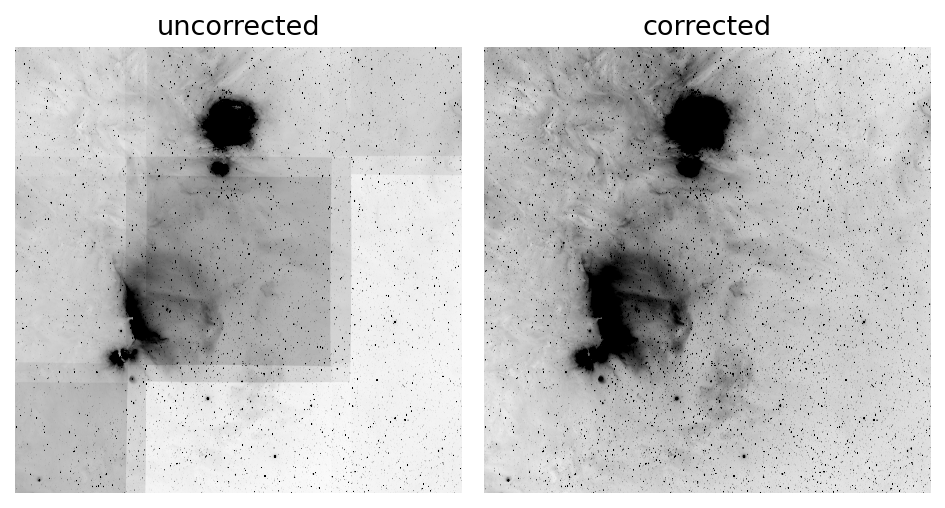}
    \caption{A comparison between directly co-adding images (left) and mosaicking after performing background corrections, as discussed in Section \ref{sec:mosaics} (right). On the left, we see prominent overlaps, due to the difference in background levels between fields. Once we correct for the varying background, the output mosaic is more uniform.}
    \label{fig:mosaic_comparison}
\end{figure}

It is possible to directly combine the reprojected images, given they share uniform coordinate information. However, due to variation in background levels between fields, this can cause the appearance of rectangular strips in the overlapping regions between adjacent fields. In standard cases, the MontagePy package offers basic background modeling features. Specifically, all overlapping areas are identified and examined, a global best-fit is determined, and each field is applied with a correcting parameter to reduce the strips. See Figure \ref{fig:mosaic_comparison} for a comparison of an uncorrected mosaic and a corrected mosaic. However, this built-in feature has a maximum number of iterations allowed for this calculation, and therefore might encounter difficulties when a large number of images are involved. Our approach is to divide larger mosaics into smaller blocks, and produce temporary mini-mosaics for each block before combining them into a single, larger mosaic. Depending on the final size of the mosaic, there could be more than one of this intermediate stage to improve the general quality and minimize the breaks between fields.

\begin{figure*}
    \centering
    \includegraphics[width=1\linewidth]{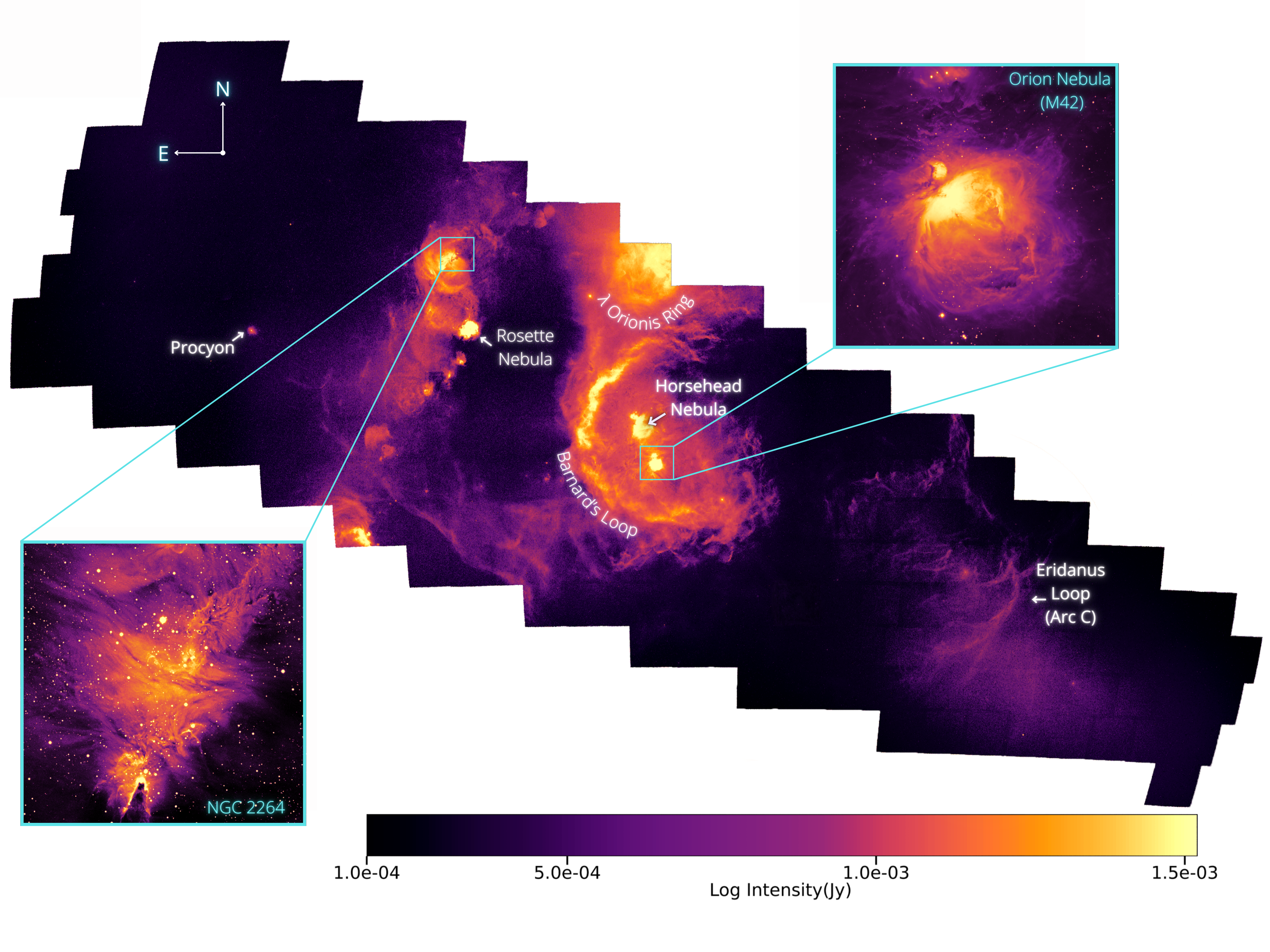}
    \caption{A mosaic of the full DR0 region, with areas of interest labeled and/or zoomed in (see Figure~\ref{fig:all_sky_map} for the coordinate range). The Orion Complex occupies the center area, which includes Barnard's Loop and the Orion nebula. In the left portion, we see the Rosette nebula and the Christmas Tree Cluster (NGC 2264) along the Galactic plane. The Eridanus Loop is also partially present. Contrast stretching has been applied to the mosaic to highlight the diffuse H$\alpha$ emission, though the magnified images of the Orion nebula (MDW field 2253) and NGC 2264 (MDW field 0368) have been specially adjusted and do not follow the color bar.} 
    \label{fig:mosaic}
\end{figure*}

Figure \ref{fig:mosaic} exhibits our final mosaic of the entire DR0 region, annotated to showcase the Orion molecular cloud complex, the Rosette and Cone nebulae, the Christmas Tree Cluster, and Arc C of the Eridanus Loop. As a side effect of the background-matching process used to mosaic the DR0 fields into a single uniform image, we found it necessary to mask the appearance of one field (field 2582). This enhancement is needed only due to the process of mosaicking, and is unrelated to the field quality. We applied contrast stretching to the mosaic, in order to accentuate the diffuse emission and filament structures.

\section{Summary \& Science Prospects}
\label{sec:summary}

The MDW H$\alpha$ Sky Survey is an ongoing imaging survey of the entire night sky in the H$\alpha$ wavelength (\textasciitilde656.3 nm). This survey holds a unique position among existing H$\alpha$ surveys in its goal of full-sky coverage at an exceptional depth and resolution, initiated and driven by a group of amateur astronomers. We believe there is ample opportunity for the survey to deepen the existing scientific knowledge on star-forming regions, supernovae remnants, and other H$\alpha$-emitting sources in the Milky Way.

This paper presents the first Data Release (DR0) of the MDW Sky Survey. Each field in the survey has an angular resolution of 3.2 arcsec/pixel, a point source depth in the 16th-17th magnitude range, and a total exposure of four hours. DR0 covers a region in Orion of area \textasciitilde3100 \textbf{${{\rm deg}^2}$} (spanning 238 fields), with Declination ranging from -30$^\circ$ to 30$^\circ$ and Right Ascension from 40$^\circ$ to 130$^\circ$. The release is composed of the following elements:
\begin{enumerate}[leftmargin=*, label=\arabic*.]
    \item A point source catalog, matched to the Pan-STARRS1 and IGAPS survey catalogs.
    \item Plate-solved, mean FITS fields (in units of counts).
    \item Plate-solved, photometrically calibrated mean FITS fields (in both star-included and star-removed varieties, in units of Jy).
    \item A mosaic of the DR0 region.
\end{enumerate}

The MDW H$\alpha$ Survey has potential applications in studying structures within the Milky Way's interstellar medium, stellar processes, and even nearby galaxies. The following are non-exhaustive examples of scientific areas where the MDW Sky Survey may be utilized:

\begin{enumerate}[leftmargin=*, label=\arabic*.]
    \item \textbf{H$\alpha$-Emitting Point Sources}: Our source catalog may be used to study and search for H$\alpha$-emitting sources, by recreating the color-color plane, described in IGAPS, with our catalog's calibrated H$\alpha$ and matched Pan-STARRS1 g, r, and i magnitudes \citep{iphas_cat_ha_emission_sources, candidate_pn_iphas_catalog}.
    \item \textbf{Study of Variable Objects}: The individual frames of each field in the MDW Sky Survey (released from DR1 onwards) provide the chance to study point source H$\alpha$-variability e.g. due to flares or chromospheric activity \citep{chromospheric_emission_f_g_k_dwarfs, ha_emission_of_m_dwarf_rotation,  chromospheric_activity_rotation_praesepe_hyades}.
    \item \textbf{Study of Extended Structures}: Images from the MDW Survey can be used to optically identify supernovae remnants \citep{supershell, igaps_sn_remnants}, stellar outflows \citep{disks_bubbles_jets_orion}, and bow-shocks \citep{bow_shocks} throughout the Milky Way.
    \item \textbf{Warm Ionized Medium (WIM)}: The WIM refers to the diffuse ionized gas in a galaxy that extends beyond stellar remnants and HII regions \citep{warm_ionized_medium}. The MDW Sky Survey can assist in studying the ionized hydrogen that makes up the WIM in the Milky Way.
    \item \textbf{High-latitude Filaments}: The MDW Sky Survey can be helpful in studying high-latitude filaments within the Milky Way. We specify high-latitude because the MDW Survey does not have velocity resolution, so filaments in the Galactic plane may be too blended to distinguish visually.
    
\end{enumerate}

Our upcoming second Data Release, DR1, will cover the entire northern sky with Dec > 0$^\circ$, totalling 2114 fields and an area of \textasciitilde20,000 ${{\rm deg}^2}$. DR1 will ultimately include the mean image of each field (similar to DR0), the 12 individual exposures that make up each field (\textasciitilde25,000 frames), and an expanded source catalog. We also intend to improve our calibration and photometry, relative to DR0. Our third Data Release, DR2, will encompass the entire sky.

\section*{In Remembrance of David R. Mittelman}

Parallels exist between the histories of the present-day MDW H$\alpha$ Sky Survey and the late 19th--early 20th century Henry Draper Catalog of Spectra \citep{draper_catalog}. Paraphrasing from the Harvard
Annals of 1892 — the Draper Catalog was made possible by the perfect storm of 3 technological breakthroughs of the day: the increased sensitivity of collodion glass plates, the use of a doublet telescope objective with an f-ratio of 5.6, capable of good focus across 10 to 25 ${\rm deg}^2$ of sky, and the incorporation of an objective prism. These capabilities allowed the collection of as many as 200 spectra on a single
glass plate. Ultimately, over the course of several decades, an unprecedented spectroscopic survey of 225,300 stars was completed. 

In the case of the MDW Sky Survey, the advent of several technologies contributed greatly to the efficacy and the practicality of an all-sky H$\alpha$ survey - an opportunity that David Mittelman realized and acted on. Reasonably priced CCD
cameras with sufficiently large format and quantum efficiencies of nearly 50\%—far exceeding even the best glass plates—became available in the early 2000. Remotely
controlled mounts such as the Software Bisque's Paramount also became available, as did the ACP software to control the planning, acquisition, calibration and execution of
nearly 4000 fields tiling the full sky. The perfection of small aperture telescopes with high index of refraction glasses and computer optimized designs, like the Astro Physics
Starfire 130, made photographs from dark sites to 20th magnitude a possibility. Telescope hosting farms, like New Mexico Skies, made access to the darkest skies a
reality. These technologies of the day, along with the vision of David Mittelman and his love of the beauty of the night sky, coalesced into the MDW H$\alpha$ Survey.

Today the technology exists to expand upon the MDW H$\alpha$ Survey.  Faster astrographs, improved filters and mounts, and low noise detectors have been developed that represent significant improvements in capability. These can be combined to uniquely extend David Mittelman's vision even further, as we contemplate future surveys of the emission-line sky. 

\section*{Acknowledgements}

Funding for the MDW H$\alpha$ Survey Project has been provided by the Michele and David Mittelman Family Foundation. David H. Sliski is the Project Director for the MDW Sky Survey and is responsible for evolving the project from an amateur sky survey into a legacy project to honor the founders. David R. Mittelman, Dennis di Cicco, and Sean Walker are founding members of the survey and made possible the acquisition and reduction of the data. The Columbia University Astronomy Department is responsible for the final data reduction, calibration, and dissemination of the survey data. 

We thank the referee for their thoughtful and constructive comments in improving this paper.  We also extend our appreciation to Lylon Sanchez Valido for her review of this manuscript and helpful edits. We are grateful to David Kipping (affiliated with Columbia University) for his support and useful discussions over the course of the survey. 

We thank the staff at New Mexico Skies Observatory for their support throughout this project, Roland and Marj Christen of Astro-Physics for their helpful discussions when selecting and modifying the survey telescopes, and Bob Denny of DC-3 Dreams for continuing to help with the automation software employed by the survey. We thank Alan Sliski for his help in producing 3D drawings of the telescopes in Solidworks, and in the construction of the third telescope for this project commissioned in 2023.

This research has made use of the VizieR catalogue access tool, CDS,
Strasbourg, France \citep{10.26093/cds/vizier}. The original description 
of the VizieR service was published in \citet{vizier}
 
The Pan-STARRS1 Surveys (PS1) and the PS1 public science archive have been made possible through contributions by the Institute for Astronomy, the University of Hawaii, the Pan-STARRS Project Office, the Max-Planck Society and its participating institutes, the Max Planck Institute for Astronomy, Heidelberg and the Max Planck Institute for Extraterrestrial Physics, Garching, The Johns Hopkins University, Durham University, the University of Edinburgh, the Queen's University Belfast, the Harvard-Smithsonian Center for Astrophysics, the Las Cumbres Observatory Global Telescope Network Incorporated, the National Central University of Taiwan, the Space Telescope Science Institute, the National Aeronautics and Space Administration under Grant No. NNX08AR22G issued through the Planetary Science Division of the NASA Science Mission Directorate, the National Science Foundation Grant No. AST-1238877, the University of Maryland, Eotvos Lorand University (ELTE), the Los Alamos National Laboratory, and the Gordon and Betty Moore Foundation.

DR0 makes use of data obtained as part of the IGAPS merger of the IPHAS and UVEX surveys (\url{www.igapsimages.org}) carried out at the Isaac Newton Telescope (INT). The INT is operated on the island of La Palma by the Isaac Newton Group in the Spanish Observatorio del Roque de los Muchachos of the Instituto de Astrofisica de Canarias. All IGAPS data were processed by the Cambridge Astronomical Survey Unit, at the Institute of Astronomy in Cambridge.

This work has made use of data from the European Space Agency (ESA) mission
{\it Gaia} (\url{https://www.cosmos.esa.int/gaia}), processed by the {\it Gaia}
Data Processing and Analysis Consortium (DPAC,
\url{https://www.cosmos.esa.int/web/gaia/dpac/consortium}). Funding for the DPAC
has been provided by national institutions, in particular the institutions
participating in the {\it Gaia} Multilateral Agreement.

This research made use of Photutils, an Astropy package for
detection and photometry of astronomical sources \citep{photutils}.

This work made use of Astropy:\footnote{\url{http://www.astropy.org}} a community-developed core Python package and an ecosystem of tools and resources for astronomy \citep{astropy}.

This work made use of Montage\footnote{\url{http://montage.ipac.caltech.edu/}}.It is funded by the National Science Foundation under Grant Number ACI-1440620, and was previously funded by the National Aeronautics and Space Administration's Earth Science Technology Office, Computation Technologies Project, under Cooperative Agreement Number NCC5-626 between NASA and the California Institute of Technology.

\

\

\textit{Software}: AstroPy \citep{astropy}, photutils \citep{photutils}, Matplotlib \citep{matplotlib}, NumPy \citep{numpy}, SciPy \citep{scipy}, Pandas \citep{pandas}, astrometry.net \citep{astrometry_net}, MontagePy \citep{montagepy}, Vizier \citep{vizier}, DC-3 Dreams \citep{acp_software}, CCDStack2 (\url{https://ccdware.com/ccdstack_overview/}), MaximDL (\url{https://diffractionlimited.com/product/maxim-dl/})


\clearpage
\appendix

\section{Telescope: Detailed Design, Tradeoffs and Modifications}
\label{appendix:a}

Before settling on the Astro-Physics 130GTX refractors, the team deliberated on multiple telescopes, including the Takahashi FSQ-106 and the Officina Stellare RH200. We initially settled on the RH200 because of its fast photographic speed.  However, the RH200's mechanical design was not fully sealed and could allow dust to enter the optical path of the filter, effecting image quality. An internal reflection would also sometimes occur when imaging near very bright stars, possibly due to the Mangin mirror used in Riccardi-Honders design. Additionally, we found it challenging to create flat-field calibration images with the RH200.

During this period of contemplation, Astro-Physics debuted its f/6.3 130GTX refractor.  This model could be modified to have a completely sealed optical path, preventing dust from obstructing the telescope. It also did not have an internal-reflection issue when imaging near bright stars, and flat-field calibration images are notably easier to obtain with refractors.

The 130GTX model can be equipped with a f/4.5 focal reducer but even with this, the 130GTX appears to have a slower photographic speed (f/4.5 compared with the RH200's f/3). On further investigation, since the RH200 has a large central obstruction and multiple reflecting surfaces, which slow down its effective photographic speed, we did not consider its speed a significant advantage over the 130GTX system that we ultimately settled on for the MDW Survey. See Table \ref{tab:telescope_comparison} for the specifications of all the telescopes we considered, and Figure \ref{fig:northern_setup} for an image of the two telescope setups in New Mexico.

\begin{figure*}
    \centering
    \includegraphics[width=0.8\textwidth]{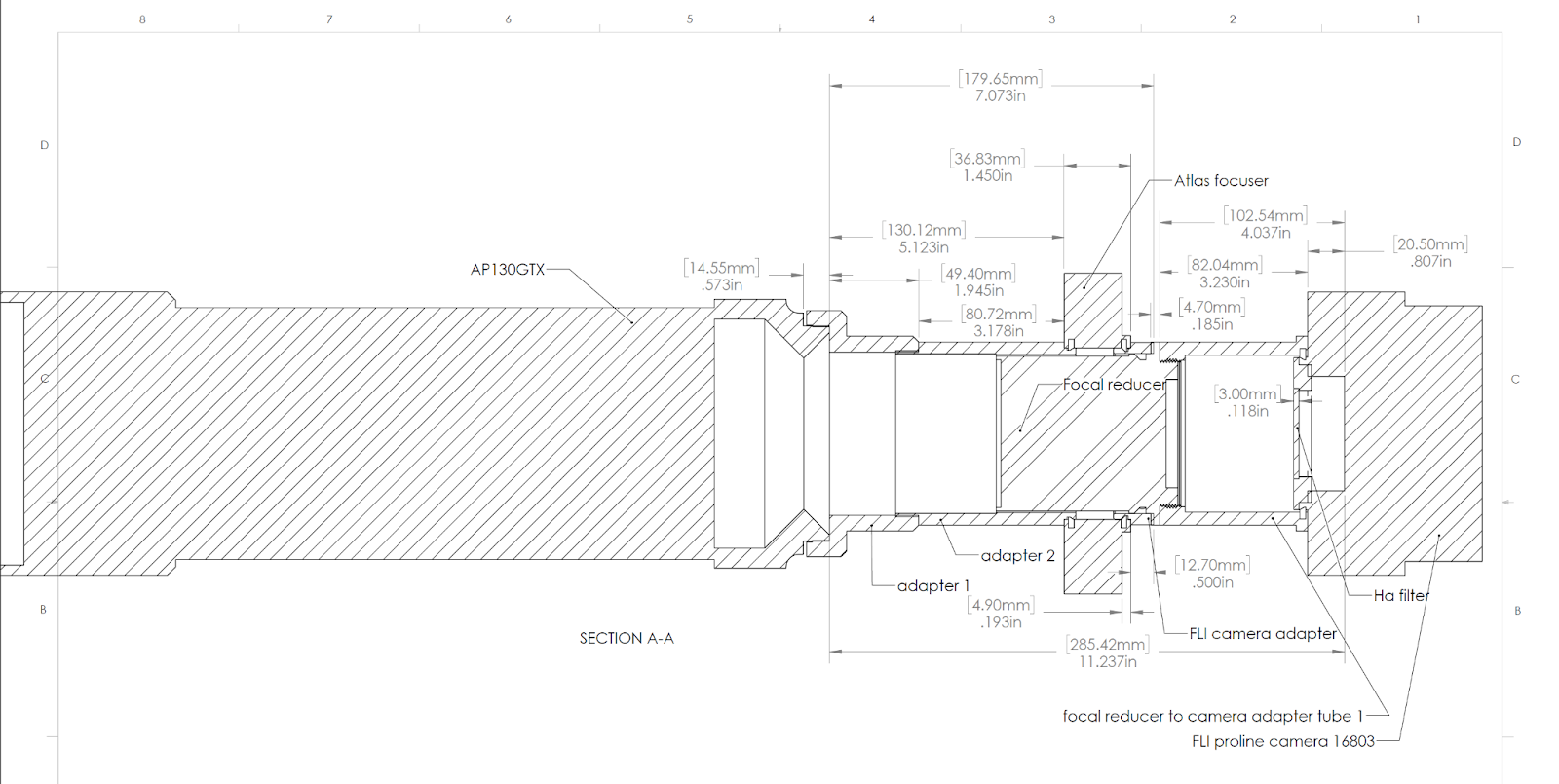}
    \caption{ \hbox{The blueprint of the MDW Survey's 130GTX refractor with modifications included.}}
    \label{fig:solidworks}
\end{figure*}

\begin{figure*}
    \centering
    \includegraphics[width=0.5\textwidth]{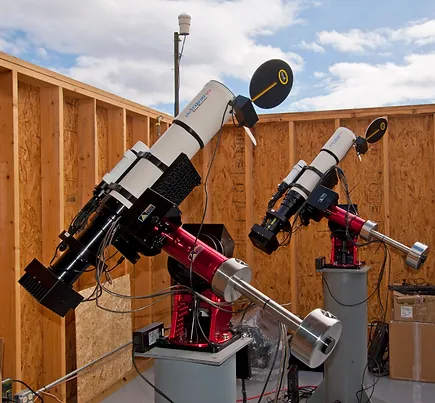}
    \caption{Pictured is our refractor set-up for the Northern half of the MDW Survey, of our two AstroPhysics 130GTX refractors with Paramount MX+ mounts. Note that we use a Paramount L-350 mount in the Southern half of our survey, carried out at ObsTech Observatario El Sauce.}
    \label{fig:northern_setup}
\end{figure*}

\begin{figure*}
    \centering
    \includegraphics[width=0.5\textwidth]{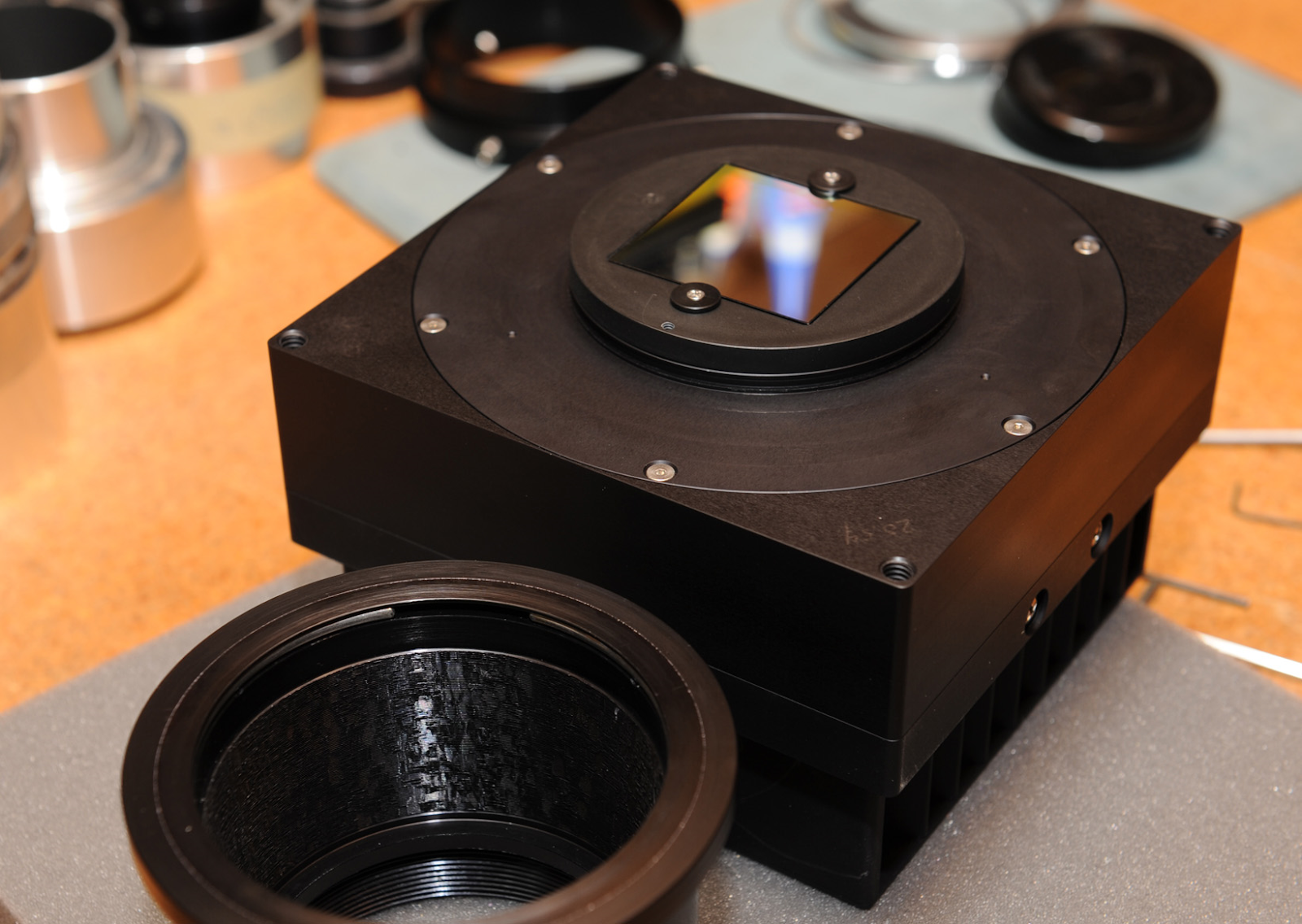}
    \caption{An FLI ProLine CCD used in the MDW Survey, with a 3nm Astrodon H$\alpha$ filter installed directly in front of it. }
    \label{fig:ccd_with_filter}
\end{figure*}

When modifying our original two 130GTX models (also see Figure \ref{fig:solidworks}), we set out to match the systems so both refractors would always maintain the same performance when imaging. We replaced the 130GTX’s focuser with non-moving adapters, eliminating a possible point of flexure in the optical train. We positioned the FLI Atlas Focuser such that the distance between the CCD and focal reducer remained fixed so that as we focused throughout the night there would be no change in optical performance. We also aimed to create a system fully sealed from dust by eliminating a conventional filter wheel, and instead permanently mounted the H$\alpha$ filter just ahead of the CCD's window (see Figure \ref{fig:ccd_with_filter}).

\begin{table*}
    \centering
    \begin{tabular}{>{\centering\arraybackslash}p{3cm} >{\centering\arraybackslash}p{3cm} >{\centering\arraybackslash}p{3cm} >{\centering\arraybackslash}p{3cm} >{\centering\arraybackslash}p{3cm}}
        \hline
        \multicolumn{1}{>{\centering\arraybackslash}p{3cm}}{\textbf{Telescope}} & 
        \multicolumn{1}{>{\centering\arraybackslash}p{3cm}}{\textbf{Design}} & 
        \multicolumn{1}{>{\centering\arraybackslash}p{3cm}}{\textbf{Aperture}} & 
        \multicolumn{1}{>{\centering\arraybackslash}p{3cm}}{\textbf{Effective Focal Length}} & 
        \multicolumn{1}{>{\centering\arraybackslash}p{3cm}}{\textbf{f-ratio}} \\
        \hline
        \multicolumn{1}{>{\centering\arraybackslash}p{3cm}}{Takahashi FSQ-106} & 
        \multicolumn{1}{>{\centering\arraybackslash}p{3cm}}{Petzval refractor} & 
        \multicolumn{1}{>{\centering\arraybackslash}p{3cm}}{106mm} & 
        \multicolumn{1}{>{\centering\arraybackslash}p{3cm}}{530mm} & 
        \multicolumn{1}{>{\centering\arraybackslash}p{3cm}}{f/5} \\
        \hline
        \multicolumn{1}{>{\centering\arraybackslash}p{3cm}}{Officina Stellare RH200} & 
        \multicolumn{1}{c}{Riccardi-Honders astrograph} & 
        \multicolumn{1}{>{\centering\arraybackslash}p{3cm}}{200mm} & 
        \multicolumn{1}{>{\centering\arraybackslash}p{3cm}}{600mm} & 
        \multicolumn{1}{>{\centering\arraybackslash}p{3cm}}{f/3 (slower in practice)} \\
        \hline
        \multicolumn{1}{>{\centering\arraybackslash}p{3cm}}{\textbf{Astro-Physics 130GTX}} & 
        \multicolumn{1}{>{\centering\arraybackslash}p{3cm}}{Apochromatic refractor} & 
        \multicolumn{1}{>{\centering\arraybackslash}p{3cm}}{130mm} & 
        \multicolumn{1}{>{\centering\arraybackslash}p{3cm}}{585mm (with focal reducer)} & 
        \multicolumn{1}{>{\centering\arraybackslash}p{3cm}}{f/4.5 (with focal reducer)} \\
        \hline
    \end{tabular}
    \caption{Comparison of the 3 telescopes considered for the MDW H$\alpha$ Sky Survey. We ultimately decided on the Astro-Physics 130GTX refractor (bolded).}
    \label{tab:telescope_comparison}
\end{table*}

\section{Survey Operations}
\label{appendix:b}

The MDW team currently utilizes the ACP software package by DC-3 Dreams to schedule observing and to integrate operation of our equipment.  Through ACP, we use TheSkyX software package to align and control our Software Bisque mounts;  MaxImDL to control the CCD's cooling and exposure time; FocusMax to measure the FWHM of stars to achieve the best focus each hour; and the PinPoint plug-in to plate-solve each field and ensure that we are pointing at the right location.

Weather conditions for observing are noted in Table \ref{tab:roof_specs}. Additionally, we reject raw exposures if the CCD's temperature does not fall within $-29$ to $-31^{\circ}$C, or if the Right Ascension (RA) and Declination (Dec) differentials are greater than 10 arcmin. Each field has an overlap of \textasciitilde11 arcmin, so the latter ensures each pointing is within the overlap error bars. We calculate the RA and Dec differentials using the equations below:

\begin{equation}
RA_{\text{differential}} = (RA_{\text{plate\_solved}} - RA_{\text{field\_center}}) \cos{Dec}
\end{equation}

\begin{equation}
Dec_{\text{differential}} = Dec_{\text{plate\_solved}} - Dec_{\text{field\_center}} \\
\end{equation}


\clearpage

 \begin{table*}
    \centering

\begin{tabular}{p{3cm}p{10cm}}
    \hline
    \multicolumn{1}{p{3cm}}{\centering \textbf{Sun}} & 
    \multicolumn{1}{p{10cm}}{\raggedright Opens 2 hours before sunset, and closes 1 hour before sunrise.} \\
    \hline
    \multicolumn{1}{p{3cm}}{\centering \textbf{Rain}} & 
    \multicolumn{1}{p{10cm}}{\raggedright Closes if 2 out of 3 rain sensors at NMS register rain, and reopens an hour after rain ends.} \\
    \hline
    \multicolumn{1}{p{3cm}}{\centering \textbf{Snow}} & 
    \multicolumn{1}{p{10cm}}{\raggedright Closes for the whole night if there is rain and temperature is less than 35$^\circ$F.} \\
    \hline
    \multicolumn{1}{p{3cm}}{\centering \textbf{Wind}} & 
    \multicolumn{1}{p{10cm}}{\raggedright Closes if windspeed > 30 mph, and reopens after 30 minutes if windspeed stays below 20 mph.} \\
    \hline
    \multicolumn{1}{p{3cm}}{\centering \textbf{Humidity}} & 
    \multicolumn{1}{p{10cm}}{\raggedright Closes if humidity > 80\%, and reopens after 30 minutes if humidity stays below 70\%.} \\
    \hline
    \multicolumn{1}{p{3cm}}{\centering \textbf{Clouds}} & 
    \multicolumn{1}{p{10cm}}{\raggedright Closes for 10 min if overcast, and reopens after 30 minutes if sky becomes clear.} \\
    \hline
\end{tabular}
    \caption{A description of when we open and close the roof at the David Mittelman Observatory to allow observing, using weather input from NMS.}
    \label{tab:roof_specs}
\end{table*}

\clearpage

\section{Catalog Columns}
\label{appendix:c}

    \begin{longtable}{|l|l|l|}
        \caption{List of columns in the DR0 point source catalog, which is matched to the Pan-STARRS1 and IGAPS survey catalogs. } 
        \label{tab:catalog_columns}
        \\
        
        \hline
        \textbf{Column} & \textbf{Type} & \textbf{Description} \\ \hhline{|=|=|=|}
        
        {MDW\_RAJ2000} & float64 & MDW J2000 Right Ascension (degrees) \\ \hline
        
        {MDW\_DEJ2000} & float64 & MDW J2000 Declination (degrees) \\ \hline
        
        {MDW\_Ha\_mag} & float64 & MDW calibrated H$\alpha$ AB magnitude \\ \hline
        
        {MDW\_Ha\_instr\_mag} & float64 & MDW instrument H$\alpha$ magnitude, calculated from MDW\_Ha\_instr\_flux \\ \hline
        
        {MDW\_Ha\_instr\_flux} & float64 & MDW instrument flux using an aperture of radius 6 arcsec (counts)\\ \hline
        
        {field} & int32 &  Field number in which the source is found in. \textsuperscript{a}\\ \hline
        
        {ps\_RAJ2000} & float64 & Pan-STARRS1 J2000 Right Ascension (degrees) \\ \hline
        
        {ps\_DEJ2000} & float64 & Pan-STARRS1 J2000 Declination (degrees) \\ \hline
        
        {ps\_objID} & int64 & Pan-STARRS1 object ID \\ \hline
        
        {ps\_g\_mag} & float64 & Pan-STARRS1 g AB magnitude \\ \hline
        
        {ps\_r\_mag} & float64 & Pan-STARRS1 r AB magnitude \\ \hline
        
        {ps\_i\_mag} & float64 & Pan-STARRS1 i AB magnitude \\ \hline
        
        {ps\_MDW\_x} & float64 &  x coordinate (pixels) in MDW image of the matched Pan-STARRS1 \textsuperscript{b} \\ \hline
        
        {ps\_MDW\_y} & float64 &  y coordinate (pixels) in MDW image of the matched Pan-STARRS1 \textsuperscript{b} \\
        \hline
        
        {in\_ig\_footprint} & bool & Whether this source is in IGAPS  \textsuperscript{c} \\ \hline
        
        {ig\_RAJ2000} & float64 &  IGAPS J2000 Right Ascension (degrees) \textsuperscript{d} \\ \hline
        
        {ig\_DEJ2000} & float64 & IGAPS J2000 Declination (degrees) \textsuperscript{d}  \\ \hline
        
        {ig\_name} & |S20 & IGAPS name. This is '-{}-' if there's no matched IGAPS source. \\ \hline
        
        {ig\_r\_mag} & float64 & IGAPS r AB magnitude \textsuperscript{d}  \\ \hline
        
        {ig\_i\_mag} & float64 & IGAPS i AB magnitude \textsuperscript{d}  \\ \hline
        
        {ig\_g\_mag} & float64 & IGAPS g AB magnitude \textsuperscript{d}  \\ \hline
        
        {ig\_Ha\_mag} & float64 & IGAPS H$\alpha$ AB magnitude. \textsuperscript{d}  \\ & & \\ & & If present, this should be similar to MDW\_Ha\_mag (Section \ref{sec:photometric_calibration}). \\ \hline
        
        {bkg\_median} & float64 &  The median background value found in the source's annulus (counts) \textsuperscript{e} \\ \hline
        
        {bkg\_total} & float64 &  The total background value of the source's annulus (counts) \textsuperscript{e}\\ \hline
        
        {bkg\_std} & float64 &  The standard deviation of the background in the source's annulus (counts) \textsuperscript{e}\\ \hline
        
       {bkg\_px\_count} & float64 & The total number of pixels contained by the annulus \textsuperscript{e}  \\ \hline
        
        {MDW\_x} & float64 & The x coordinate where the MDW source is found in the field image \\ \hline
        
        {MDW\_y} & float64 & The y coordinate where the MDW source is found in the field image \\ \hline
        
        {sep2d} & float64 &  The 2D separation distance between the MDW source and the matched Pan-STARRS1 source \textsuperscript{b} \\ \hline

    \end{longtable}

    \begin{tablenotes}
            \small 
            \item{$^\text{a}$ \footnotesize In practice, a source can be found in two fields because of observing overlaps (Section \ref{sec:implementation}). We intend to refine the catalog for DR1 to account for this possibility.}

            \item{$^\text{b}$ \footnotesize When there's a matched IGAPS source, the corresponding  value from the IGAPS matching is close to the matched Pan-STARRS1. As a result, we only keep the Pan-STARRS1 column in the catalog for conciseness. }

            \item{$^\text{c}$ \footnotesize If this is true, it doesn't necessarily mean we were able to match and populate IGAPS data for this source a.k.a \texttt{(in\_ig\_footprint \& ig\_name =='-{}-')} can be true.}

            \item{$^\text{d}$ \footnotesize If there's no matched IGAPS source, this value is -999.}

            \item{$^\text{e}$ \footnotesize In our aperture photometry, we use an annulus of width 6 arcsec (or \textasciitilde2 pixels), with an inner radius at 8 arcsec and outer radius at 14 arcsec.}
        \end{tablenotes}
\clearpage

\bibliographystyle{aasjournal}
\bibliography{dr0_citations} 

\begin{thebibliography}{}
\expandafter\ifx\csname natexlab\endcsname\relax\def\natexlab#1{#1}\fi
\providecommand{\url}[1]{\href{#1}{#1}}
\providecommand{\dodoi}[1]{doi:~\href{http://doi.org/#1}{\nolinkurl{#1}}}
\providecommand{\doeprint}[1]{\href{http://ascl.net/#1}{\nolinkurl{http://ascl.net/#1}}}
\providecommand{\doarXiv}[1]{\href{https://arxiv.org/abs/#1}{\nolinkurl{https://arxiv.org/abs/#1}}}

\bibitem[{{Adams} {et~al.}(2011){Adams}, {Uson}, {Hill}, \& {MacQueen}}]{adams11}
{Adams}, J.~J., {Uson}, J.~M., {Hill}, G.~J., \& {MacQueen}, P.~J. 2011, \apj, 728, 107, \dodoi{10.1088/0004-637X/728/2/107}

\bibitem[{{Aftab} {et~al.}(2024){Aftab}, {Homa}, {Holm-Hansen}, {Zhang}, {Schiminovich}, {Putman}, {Mittelman}, {di Cicco}, \& {Walker}}]{noor_aas_iposter}
{Aftab}, N., {Homa}, J., {Holm-Hansen}, C., {et~al.} 2024, in American Astronomical Society Meeting Abstracts, Vol. 243, American Astronomical Society Meeting Abstracts, 361.02

\bibitem[{{Astropy Collaboration} {et~al.}(2022){Astropy Collaboration}, {Price-Whelan}, {Lim}, {Earl}, {Starkman}, {Bradley}, {Shupe}, {Patil}, {Corrales}, {Brasseur}, {N{\"o}the}, {Donath}, {Tollerud}, {Morris}, {Ginsburg}, {Vaher}, {Weaver}, {Tocknell}, {Jamieson}, {van Kerkwijk}, {Robitaille}, {Merry}, {Bachetti}, {G{\"u}nther}, {Aldcroft}, {Alvarado-Montes}, {Archibald}, {B{\'o}di}, {Bapat}, {Barentsen}, {Baz{\'a}n}, {Biswas}, {Boquien}, {Burke}, {Cara}, {Cara}, {Conroy}, {Conseil}, {Craig}, {Cross}, {Cruz}, {D'Eugenio}, {Dencheva}, {Devillepoix}, {Dietrich}, {Eigenbrot}, {Erben}, {Ferreira}, {Foreman-Mackey}, {Fox}, {Freij}, {Garg}, {Geda}, {Glattly}, {Gondhalekar}, {Gordon}, {Grant}, {Greenfield}, {Groener}, {Guest}, {Gurovich}, {Handberg}, {Hart}, {Hatfield-Dodds}, {Homeier}, {Hosseinzadeh}, {Jenness}, {Jones}, {Joseph}, {Kalmbach}, {Karamehmetoglu}, {Ka{\l}uszy{\'n}ski}, {Kelley}, {Kern}, {Kerzendorf}, {Koch}, {Kulumani}, {Lee}, {Ly}, {Ma}, {MacBride}, {Maljaars}, {Muna}, {Murphy}, {Norman},
  {O'Steen}, {Oman}, {Pacifici}, {Pascual}, {Pascual-Granado}, {Patil}, {Perren}, {Pickering}, {Rastogi}, {Roulston}, {Ryan}, {Rykoff}, {Sabater}, {Sakurikar}, {Salgado}, {Sanghi}, {Saunders}, {Savchenko}, {Schwardt}, {Seifert-Eckert}, {Shih}, {Jain}, {Shukla}, {Sick}, {Simpson}, {Singanamalla}, {Singer}, {Singhal}, {Sinha}, {Sip{\H{o}}cz}, {Spitler}, {Stansby}, {Streicher}, {{\v{S}}umak}, {Swinbank}, {Taranu}, {Tewary}, {Tremblay}, {de Val-Borro}, {Van Kooten}, {Vasovi{\'c}}, {Verma}, {de Miranda Cardoso}, {Williams}, {Wilson}, {Winkel}, {Wood-Vasey}, {Xue}, {Yoachim}, {Zhang}, {Zonca}, \& {Astropy Project Contributors}}]{astropy}
{Astropy Collaboration}, {Price-Whelan}, A.~M., {Lim}, P.~L., {et~al.} 2022, \apj, 935, 167, \dodoi{10.3847/1538-4357/ac7c74}

\bibitem[{{Bally} {et~al.}(2000){Bally}, {O'Dell}, \& {McCaughrean}}]{disks_bubbles_jets_orion}
{Bally}, J., {O'Dell}, C.~R., \& {McCaughrean}, M.~J. 2000, \aj, 119, 2919, \dodoi{10.1086/301385}

\bibitem[{{Barentsen} {et~al.}(2014){Barentsen}, {Farnhill}, {Drew}, {Gonz{\'a}lez-Solares}, {Greimel}, {Irwin}, {Miszalski}, {Ruhland}, {Groot}, {Mampaso}, {Sale}, {Henden}, {Aungwerojwit}, {Barlow}, {Carter}, {Corradi}, {Drake}, {Eisl{\"o}ffel}, {Fabregat}, {G{\"a}nsicke}, {Gentile Fusillo}, {Greiss}, {Hales}, {Hodgkin}, {Huckvale}, {Irwin}, {King}, {Knigge}, {Kupfer}, {Lagadec}, {Lennon}, {Lewis}, {Mohr-Smith}, {Morris}, {Naylor}, {Parker}, {Phillipps}, {Pyrzas}, {Raddi}, {Roelofs}, {Rodr{\'\i}guez-Gil}, {Sabin}, {Scaringi}, {Steeghs}, {Suso}, {Tata}, {Unruh}, {van Roestel}, {Viironen}, {Vink}, {Walton}, {Wright}, \& {Zijlstra}}]{IPHAS}
{Barentsen}, G., {Farnhill}, H.~J., {Drew}, J.~E., {et~al.} 2014, \mnras, 444, 3230, \dodoi{10.1093/mnras/stu1651}

\bibitem[{{Bracco} {et~al.}(2020){Bracco}, {Benjamin}, {Alves}, {Lehmann}, {Boulanger}, {Montier}, {Mittelman}, {di Cicco}, \& {Walker}}]{ursa_major_arc}
{Bracco}, A., {Benjamin}, R.~A., {Alves}, M.~I.~R., {et~al.} 2020, \aap, 636, L8, \dodoi{10.1051/0004-6361/202037975}

\bibitem[{Bradley {et~al.}(2023)Bradley, Sip{\H o}cz, Robitaille, Tollerud, Vin{\'{\i}}cius, Deil, Barbary, Wilson, Busko, Donath, G{\"u}nther, Cara, Lim, Me{\ss}linger, Conseil, Bostroem, Droettboom, Bray, Bratholm, Barentsen, Craig, Rathi, Pascual, Perren, Georgiev, de~Val-Borro, Kerzendorf, Bach, Quint, \& Souchereau}]{photutils}
Bradley, L., Sip{\H o}cz, B., Robitaille, T., {et~al.} 2023, astropy/photutils: 1.8.0, 1.8.0,  Zenodo, \dodoi{10.5281/zenodo.7946442}

\bibitem[{{Chambers} {et~al.}(2016){Chambers}, {Magnier}, {Metcalfe}, {Flewelling}, {Huber}, {Waters}, {Denneau}, {Draper}, {Farrow}, {Finkbeiner}, {Holmberg}, {Koppenhoefer}, {Price}, {Rest}, {Saglia}, {Schlafly}, {Smartt}, {Sweeney}, {Wainscoat}, {Burgett}, {Chastel}, {Grav}, {Heasley}, {Hodapp}, {Jedicke}, {Kaiser}, {Kudritzki}, {Luppino}, {Lupton}, {Monet}, {Morgan}, {Onaka}, {Shiao}, {Stubbs}, {Tonry}, {White}, {Ba{\~n}ados}, {Bell}, {Bender}, {Bernard}, {Boegner}, {Boffi}, {Botticella}, {Calamida}, {Casertano}, {Chen}, {Chen}, {Cole}, {Deacon}, {Frenk}, {Fitzsimmons}, {Gezari}, {Gibbs}, {Goessl}, {Goggia}, {Gourgue}, {Goldman}, {Grant}, {Grebel}, {Hambly}, {Hasinger}, {Heavens}, {Heckman}, {Henderson}, {Henning}, {Holman}, {Hopp}, {Ip}, {Isani}, {Jackson}, {Keyes}, {Koekemoer}, {Kotak}, {Le}, {Liska}, {Long}, {Lucey}, {Liu}, {Martin}, {Masci}, {McLean}, {Mindel}, {Misra}, {Morganson}, {Murphy}, {Obaika}, {Narayan}, {Nieto-Santisteban}, {Norberg}, {Peacock}, {Pier}, {Postman}, {Primak}, {Rae}, {Rai},
  {Riess}, {Riffeser}, {Rix}, {R{\"o}ser}, {Russel}, {Rutz}, {Schilbach}, {Schultz}, {Scolnic}, {Strolger}, {Szalay}, {Seitz}, {Small}, {Smith}, {Soderblom}, {Taylor}, {Thomson}, {Taylor}, {Thakar}, {Thiel}, {Thilker}, {Unger}, {Urata}, {Valenti}, {Wagner}, {Walder}, {Walter}, {Watters}, {Werner}, {Wood-Vasey}, \& {Wyse}}]{panstarrs_surveys}
{Chambers}, K.~C., {Magnier}, E.~A., {Metcalfe}, N., {et~al.} 2016, arXiv e-prints, arXiv:1612.05560, \dodoi{10.48550/arXiv.1612.05560}

\bibitem[{{Churchwell}(2002)}]{churchill02}
{Churchwell}, E. 2002, \araa, 40, 27, \dodoi{10.1146/annurev.astro.40.060401.093845}

\bibitem[{{Cordes} {et~al.}(1993){Cordes}, {Romani}, \& {Lundgren}}]{bow_shocks}
{Cordes}, J.~M., {Romani}, R.~W., \& {Lundgren}, S.~C. 1993, \nat, 362, 133, \dodoi{10.1038/362133a0}

\bibitem[{{Courtes}(1951{\natexlab{a}})}]{Courtes1951a}
{Courtes}, G. 1951{\natexlab{a}}, Academie des Sciences Paris Comptes Rendus Serie B Sciences Physiques, 232, 795

\bibitem[{{Courtes}(1951{\natexlab{b}})}]{Courtes1951b}
---. 1951{\natexlab{b}}, Academie des Sciences Paris Comptes Rendus Serie B Sciences Physiques, 232, 1283

\bibitem[{{Dennison} {et~al.}(1998{\natexlab{a}}){Dennison}, {Simonetti}, \& {Topasna}}]{dennison98}
{Dennison}, B., {Simonetti}, J.~H., \& {Topasna}, G.~A. 1998{\natexlab{a}}, \pasa, 15, 147, \dodoi{10.1071/AS98147}

\bibitem[{{Dennison} {et~al.}(1998{\natexlab{b}}){Dennison}, {Simonetti}, \& {Topasna}}]{VTSS}
---. 1998{\natexlab{b}}, \pasa, 15, 147, \dodoi{10.1071/AS98147}

\bibitem[{{Denny}(2011)}]{acp_software}
{Denny}, R. 2011, in Telescopes from Afar, ed. S.~{Gajadhar}, J.~{Walawender}, R.~{Genet}, C.~{Veillet}, A.~{Adamson}, J.~{Martinez}, J.~{Melnik}, T.~{Jenness}, \& N.~{Manset}, 47

\bibitem[{di~Cicco(2019)}]{st_article}
di~Cicco, D. 2019, Sky \& Telescope, 138, 21–27

\bibitem[{{Drechsler} {et~al.}(2023){Drechsler}, {Strottner}, {Sainty}, {Fesen}, {Kimeswenger}, {Shull}, {Falls}, {Vergnes}, {Martino}, \& {Walker}}]{m31_oiii_emission}
{Drechsler}, M., {Strottner}, X., {Sainty}, Y., {et~al.} 2023, Research Notes of the American Astronomical Society, 7, 1, \dodoi{10.3847/2515-5172/acaf7e}

\bibitem[{{Drew} {et~al.}(2014){Drew}, {Gonzalez-Solares}, {Greimel}, {Irwin}, {K{\"u}pc{\"u} Yoldas}, {Lewis}, {Barentsen}, {Eisl{\"o}ffel}, {Farnhill}, {Martin}, {Walsh}, {Walton}, {Mohr-Smith}, {Raddi}, {Sale}, {Wright}, {Groot}, {Barlow}, {Corradi}, {Drake}, {Fabregat}, {Frew}, {G{\"a}nsicke}, {Knigge}, {Mampaso}, {Morris}, {Naylor}, {Parker}, {Phillipps}, {Ruhland}, {Steeghs}, {Unruh}, {Vink}, {Wesson}, \& {Zijlstra}}]{VPHAS}
{Drew}, J.~E., {Gonzalez-Solares}, E., {Greimel}, R., {et~al.} 2014, \mnras, 440, 2036, \dodoi{10.1093/mnras/stu394}

\bibitem[{{Fesen} {et~al.}(2020){Fesen}, {Weil}, {Raymond}, {Huet}, {Rusterholz}, {di Cicco}, {Mittelman}, {Walker}, {Drechsler}, \& {Faworski}}]{supernova_remnant_cepheus}
{Fesen}, R.~A., {Weil}, K.~E., {Raymond}, J.~C., {et~al.} 2020, \mnras, 498, 5194, \dodoi{10.1093/mnras/staa2765}

\bibitem[{{Fesen} {et~al.}(2021){Fesen}, {Drechsler}, {Weil}, {Strottner}, {Raymond}, {Rupert}, {Milisavljevic}, {Subrayan}, {di Cicco}, {Walker}, {Mittelman}, \& {Ludgate}}]{uv_optical_emission_high_latitude_snr}
{Fesen}, R.~A., {Drechsler}, M., {Weil}, K.~E., {et~al.} 2021, \apj, 920, 90, \dodoi{10.3847/1538-4357/ac0ada}

\bibitem[{Fesen {et~al.}(2024)Fesen, Drechsler, Strottner, Falls, Sainty, Martino, Galli, Ludgate, Blauensteiner, Reich, Walker, di~Cicco, Mittelman, Morgan, Kaeouach, Rupert, \& Benkhaldoun}]{fesen2024deep}
Fesen, R.~A., Drechsler, M., Strottner, X., {et~al.} 2024, Deep Optical Emission-Line Images of Nine Known and Three New Galactic Supernova Remnants.
\newblock \doarXiv{2403.00317}

\bibitem[{{Fumagalli} {et~al.}(2017){Fumagalli}, {Haardt}, {Theuns}, {Morris}, {Cantalupo}, {Madau}, \& {Fossati}}]{fumagalli17}
{Fumagalli}, M., {Haardt}, F., {Theuns}, T., {et~al.} 2017, \mnras, 467, 4802, \dodoi{10.1093/mnras/stx398}

\bibitem[{{Gaia Collaboration} {et~al.}(2016){Gaia Collaboration}, {Prusti}, {de Bruijne}, {Brown}, {Vallenari}, {Babusiaux}, {Bailer-Jones}, {Bastian}, {Biermann}, {Evans}, {Eyer}, {Jansen}, {Jordi}, {Klioner}, {Lammers}, {Lindegren}, {Luri}, {Mignard}, {Milligan}, {Panem}, {Poinsignon}, {Pourbaix}, {Randich}, {Sarri}, {Sartoretti}, {Siddiqui}, {Soubiran}, {Valette}, {van Leeuwen}, {Walton}, {Aerts}, {Arenou}, {Cropper}, {Drimmel}, {H{\o}g}, {Katz}, {Lattanzi}, {O'Mullane}, {Grebel}, {Holland}, {Huc}, {Passot}, {Bramante}, {Cacciari}, {Casta{\~n}eda}, {Chaoul}, {Cheek}, {De Angeli}, {Fabricius}, {Guerra}, {Hern{\'a}ndez}, {Jean-Antoine-Piccolo}, {Masana}, {Messineo}, {Mowlavi}, {Nienartowicz}, {Ord{\'o}{\~n}ez-Blanco}, {Panuzzo}, {Portell}, {Richards}, {Riello}, {Seabroke}, {Tanga}, {Th{\'e}venin}, {Torra}, {Els}, {Gracia-Abril}, {Comoretto}, {Garcia-Reinaldos}, {Lock}, {Mercier}, {Altmann}, {Andrae}, {Astraatmadja}, {Bellas-Velidis}, {Benson}, {Berthier}, {Blomme}, {Busso}, {Carry}, {Cellino}, {Clementini},
  {Cowell}, {Creevey}, {Cuypers}, {Davidson}, {De Ridder}, {de Torres}, {Delchambre}, {Dell'Oro}, {Ducourant}, {Fr{\'e}mat}, {Garc{\'\i}a-Torres}, {Gosset}, {Halbwachs}, {Hambly}, {Harrison}, {Hauser}, {Hestroffer}, {Hodgkin}, {Huckle}, {Hutton}, {Jasniewicz}, {Jordan}, {Kontizas}, {Korn}, {Lanzafame}, {Manteiga}, {Moitinho}, {Muinonen}, {Osinde}, {Pancino}, {Pauwels}, {Petit}, {Recio-Blanco}, {Robin}, {Sarro}, {Siopis}, {Smith}, {Smith}, {Sozzetti}, {Thuillot}, {van Reeven}, {Viala}, {Abbas}, {Abreu Aramburu}, {Accart}, {Aguado}, {Allan}, {Allasia}, {Altavilla}, {{\'A}lvarez}, {Alves}, {Anderson}, {Andrei}, {Anglada Varela}, {Antiche}, {Antoja}, {Ant{\'o}n}, {Arcay}, {Atzei}, {Ayache}, {Bach}, {Baker}, {Balaguer-N{\'u}{\~n}ez}, {Barache}, {Barata}, {Barbier}, {Barblan}, {Baroni}, {Barrado y Navascu{\'e}s}, {Barros}, {Barstow}, {Becciani}, {Bellazzini}, {Bellei}, {Bello Garc{\'\i}a}, {Belokurov}, {Bendjoya}, {Berihuete}, {Bianchi}, {Bienaym{\'e}}, {Billebaud}, {Blagorodnova}, {Blanco-Cuaresma}, {Boch},
  {Bombrun}, {Borrachero}, {Bouquillon}, {Bourda}, {Bouy}, {Bragaglia}, {Breddels}, {Brouillet}, {Br{\"u}semeister}, {Bucciarelli}, {Budnik}, {Burgess}, {Burgon}, {Burlacu}, {Busonero}, {Buzzi}, {Caffau}, {Cambras}, {Campbell}, {Cancelliere}, {Cantat-Gaudin}, {Carlucci}, {Carrasco}, {Castellani}, {Charlot}, {Charnas}, {Charvet}, {Chassat}, {Chiavassa}, {Clotet}, {Cocozza}, {Collins}, {Collins}, {Costigan}, {Crifo}, {Cross}, {Crosta}, {Crowley}, {Dafonte}, {Damerdji}, {Dapergolas}, {David}, {David}, {De Cat}, {de Felice}, {de Laverny}, {De Luise}, {De March}, {de Martino}, {de Souza}, {Debosscher}, {del Pozo}, {Delbo}, {Delgado}, {Delgado}, {di Marco}, {Di Matteo}, {Diakite}, {Distefano}, {Dolding}, {Dos Anjos}, {Drazinos}, {Dur{\'a}n}, {Dzigan}, {Ecale}, {Edvardsson}, {Enke}, {Erdmann}, {Escolar}, {Espina}, {Evans}, {Eynard Bontemps}, {Fabre}, {Fabrizio}, {Faigler}, {Falc{\~a}o}, {Farr{\`a}s Casas}, {Faye}, {Federici}, {Fedorets}, {Fern{\'a}ndez-Hern{\'a}ndez}, {Fernique}, {Fienga}, {Figueras}, {Filippi},
  {Findeisen}, {Fonti}, {Fouesneau}, {Fraile}, {Fraser}, {Fuchs}, {Furnell}, {Gai}, {Galleti}, {Galluccio}, {Garabato}, {Garc{\'\i}a-Sedano}, {Gar{\'e}}, {Garofalo}, {Garralda}, {Gavras}, {Gerssen}, {Geyer}, {Gilmore}, {Girona}, {Giuffrida}, {Gomes}, {Gonz{\'a}lez-Marcos}, {Gonz{\'a}lez-N{\'u}{\~n}ez}, {Gonz{\'a}lez-Vidal}, {Granvik}, {Guerrier}, {Guillout}, {Guiraud}, {G{\'u}rpide}, {Guti{\'e}rrez-S{\'a}nchez}, {Guy}, {Haigron}, {Hatzidimitriou}, {Haywood}, {Heiter}, {Helmi}, {Hobbs}, {Hofmann}, {Holl}, {Holland}, {Hunt}, {Hypki}, {Icardi}, {Irwin}, {Jevardat de Fombelle}, {Jofr{\'e}}, {Jonker}, {Jorissen}, {Julbe}, {Karampelas}, {Kochoska}, {Kohley}, {Kolenberg}, {Kontizas}, {Koposov}, {Kordopatis}, {Koubsky}, {Kowalczyk}, {Krone-Martins}, {Kudryashova}, {Kull}, {Bachchan}, {Lacoste-Seris}, {Lanza}, {Lavigne}, {Le Poncin-Lafitte}, {Lebreton}, {Lebzelter}, {Leccia}, {Leclerc}, {Lecoeur-Taibi}, {Lemaitre}, {Lenhardt}, {Leroux}, {Liao}, {Licata}, {Lindstr{\o}m}, {Lister}, {Livanou}, {Lobel}, {L{\"o}ffler},
  {L{\'o}pez}, {Lopez-Lozano}, {Lorenz}, {Loureiro}, {MacDonald}, {Magalh{\~a}es Fernandes}, {Managau}, {Mann}, {Mantelet}, {Marchal}, {Marchant}, {Marconi}, {Marie}, {Marinoni}, {Marrese}, {Marschalk{\'o}}, {Marshall}, {Mart{\'\i}n-Fleitas}, {Martino}, {Mary}, {Matijevi{\v{c}}}, {Mazeh}, {McMillan}, {Messina}, {Mestre}, {Michalik}, {Millar}, {Miranda}, {Molina}, {Molinaro}, {Molinaro}, {Moln{\'a}r}, {Moniez}, {Montegriffo}, {Monteiro}, {Mor}, {Mora}, {Morbidelli}, {Morel}, {Morgenthaler}, {Morley}, {Morris}, {Mulone}, {Muraveva}, {Musella}, {Narbonne}, {Nelemans}, {Nicastro}, {Noval}, {Ord{\'e}novic}, {Ordieres-Mer{\'e}}, {Osborne}, {Pagani}, {Pagano}, {Pailler}, {Palacin}, {Palaversa}, {Parsons}, {Paulsen}, {Pecoraro}, {Pedrosa}, {Pentik{\"a}inen}, {Pereira}, {Pichon}, {Piersimoni}, {Pineau}, {Plachy}, {Plum}, {Poujoulet}, {Pr{\v{s}}a}, {Pulone}, {Ragaini}, {Rago}, {Rambaux}, {Ramos-Lerate}, {Ranalli}, {Rauw}, {Read}, {Regibo}, {Renk}, {Reyl{\'e}}, {Ribeiro}, {Rimoldini}, {Ripepi}, {Riva}, {Rixon},
  {Roelens}, {Romero-G{\'o}mez}, {Rowell}, {Royer}, {Rudolph}, {Ruiz-Dern}, {Sadowski}, {Sagrist{\`a} Sell{\'e}s}, {Sahlmann}, {Salgado}, {Salguero}, {Sarasso}, {Savietto}, {Schnorhk}, {Schultheis}, {Sciacca}, {Segol}, {Segovia}, {Segransan}, {Serpell}, {Shih}, {Smareglia}, {Smart}, {Smith}, {Solano}, {Solitro}, {Sordo}, {Soria Nieto}, {Souchay}, {Spagna}, {Spoto}, {Stampa}, {Steele}, {Steidelm{\"u}ller}, {Stephenson}, {Stoev}, {Suess}, {S{\"u}veges}, {Surdej}, {Szabados}, {Szegedi-Elek}, {Tapiador}, {Taris}, {Tauran}, {Taylor}, {Teixeira}, {Terrett}, {Tingley}, {Trager}, {Turon}, {Ulla}, {Utrilla}, {Valentini}, {van Elteren}, {Van Hemelryck}, {van Leeuwen}, {Varadi}, {Vecchiato}, {Veljanoski}, {Via}, {Vicente}, {Vogt}, {Voss}, {Votruba}, {Voutsinas}, {Walmsley}, {Weiler}, {Weingrill}, {Werner}, {Wevers}, {Whitehead}, {Wyrzykowski}, {Yoldas}, {{\v{Z}}erjal}, {Zucker}, {Zurbach}, {Zwitter}, {Alecu}, {Allen}, {Allende Prieto}, {Amorim}, {Anglada-Escud{\'e}}, {Arsenijevic}, {Azaz}, {Balm}, {Beck}, {Bernstein},
  {Bigot}, {Bijaoui}, {Blasco}, {Bonfigli}, {Bono}, {Boudreault}, {Bressan}, {Brown}, {Brunet}, {Bunclark}, {Buonanno}, {Butkevich}, {Carret}, {Carrion}, {Chemin}, {Ch{\'e}reau}, {Corcione}, {Darmigny}, {de Boer}, {de Teodoro}, {de Zeeuw}, {Delle Luche}, {Domingues}, {Dubath}, {Fodor}, {Fr{\'e}zouls}, {Fries}, {Fustes}, {Fyfe}, {Gallardo}, {Gallegos}, {Gardiol}, {Gebran}, {Gomboc}, {G{\'o}mez}, {Grux}, {Gueguen}, {Heyrovsky}, {Hoar}, {Iannicola}, {Isasi Parache}, {Janotto}, {Joliet}, {Jonckheere}, {Keil}, {Kim}, {Klagyivik}, {Klar}, {Knude}, {Kochukhov}, {Kolka}, {Kos}, {Kutka}, {Lainey}, {LeBouquin}, {Liu}, {Loreggia}, {Makarov}, {Marseille}, {Martayan}, {Martinez-Rubi}, {Massart}, {Meynadier}, {Mignot}, {Munari}, {Nguyen}, {Nordlander}, {Ocvirk}, {O'Flaherty}, {Olias Sanz}, {Ortiz}, {Osorio}, {Oszkiewicz}, {Ouzounis}, {Palmer}, {Park}, {Pasquato}, {Peltzer}, {Peralta}, {P{\'e}turaud}, {Pieniluoma}, {Pigozzi}, {Poels}, {Prat}, {Prod'homme}, {Raison}, {Rebordao}, {Risquez}, {Rocca-Volmerange}, {Rosen},
  {Ruiz-Fuertes}, {Russo}, {Sembay}, {Serraller Vizcaino}, {Short}, {Siebert}, {Silva}, {Sinachopoulos}, {Slezak}, {Soffel}, {Sosnowska}, {Strai{\v{z}}ys}, {ter Linden}, {Terrell}, {Theil}, {Tiede}, {Troisi}, {Tsalmantza}, {Tur}, {Vaccari}, {Vachier}, {Valles}, {Van Hamme}, {Veltz}, {Virtanen}, {Wallut}, {Wichmann}, {Wilkinson}, {Ziaeepour}, \& {Zschocke}}]{gaia_mission}
{Gaia Collaboration}, {Prusti}, T., {de Bruijne}, J.~H.~J., {et~al.} 2016, \aap, 595, A1, \dodoi{10.1051/0004-6361/201629272}

\bibitem[{{Gaia Collaboration} {et~al.}(2023){Gaia Collaboration}, {Vallenari}, {Brown}, {Prusti}, {de Bruijne}, {Arenou}, {Babusiaux}, {Biermann}, {Creevey}, {Ducourant}, {Evans}, {Eyer}, {Guerra}, {Hutton}, {Jordi}, {Klioner}, {Lammers}, {Lindegren}, {Luri}, {Mignard}, {Panem}, {Pourbaix}, {Randich}, {Sartoretti}, {Soubiran}, {Tanga}, {Walton}, {Bailer-Jones}, {Bastian}, {Drimmel}, {Jansen}, {Katz}, {Lattanzi}, {van Leeuwen}, {Bakker}, {Cacciari}, {Casta{\~n}eda}, {De Angeli}, {Fabricius}, {Fouesneau}, {Fr{\'e}mat}, {Galluccio}, {Guerrier}, {Heiter}, {Masana}, {Messineo}, {Mowlavi}, {Nicolas}, {Nienartowicz}, {Pailler}, {Panuzzo}, {Riclet}, {Roux}, {Seabroke}, {Sordo}, {Th{\'e}venin}, {Gracia-Abril}, {Portell}, {Teyssier}, {Altmann}, {Andrae}, {Audard}, {Bellas-Velidis}, {Benson}, {Berthier}, {Blomme}, {Burgess}, {Busonero}, {Busso}, {C{\'a}novas}, {Carry}, {Cellino}, {Cheek}, {Clementini}, {Damerdji}, {Davidson}, {de Teodoro}, {Nu{\~n}ez Campos}, {Delchambre}, {Dell'Oro}, {Esquej},
  {Fern{\'a}ndez-Hern{\'a}ndez}, {Fraile}, {Garabato}, {Garc{\'\i}a-Lario}, {Gosset}, {Haigron}, {Halbwachs}, {Hambly}, {Harrison}, {Hern{\'a}ndez}, {Hestroffer}, {Hodgkin}, {Holl}, {Jan{\ss}en}, {Jevardat de Fombelle}, {Jordan}, {Krone-Martins}, {Lanzafame}, {L{\"o}ffler}, {Marchal}, {Marrese}, {Moitinho}, {Muinonen}, {Osborne}, {Pancino}, {Pauwels}, {Recio-Blanco}, {Reyl{\'e}}, {Riello}, {Rimoldini}, {Roegiers}, {Rybizki}, {Sarro}, {Siopis}, {Smith}, {Sozzetti}, {Utrilla}, {van Leeuwen}, {Abbas}, {{\'A}brah{\'a}m}, {Abreu Aramburu}, {Aerts}, {Aguado}, {Ajaj}, {Aldea-Montero}, {Altavilla}, {{\'A}lvarez}, {Alves}, {Anders}, {Anderson}, {Anglada Varela}, {Antoja}, {Baines}, {Baker}, {Balaguer-N{\'u}{\~n}ez}, {Balbinot}, {Balog}, {Barache}, {Barbato}, {Barros}, {Barstow}, {Bartolom{\'e}}, {Bassilana}, {Bauchet}, {Becciani}, {Bellazzini}, {Berihuete}, {Bernet}, {Bertone}, {Bianchi}, {Binnenfeld}, {Blanco-Cuaresma}, {Blazere}, {Boch}, {Bombrun}, {Bossini}, {Bouquillon}, {Bragaglia}, {Bramante}, {Breedt},
  {Bressan}, {Brouillet}, {Brugaletta}, {Bucciarelli}, {Burlacu}, {Butkevich}, {Buzzi}, {Caffau}, {Cancelliere}, {Cantat-Gaudin}, {Carballo}, {Carlucci}, {Carnerero}, {Carrasco}, {Casamiquela}, {Castellani}, {Castro-Ginard}, {Chaoul}, {Charlot}, {Chemin}, {Chiaramida}, {Chiavassa}, {Chornay}, {Comoretto}, {Contursi}, {Cooper}, {Cornez}, {Cowell}, {Crifo}, {Cropper}, {Crosta}, {Crowley}, {Dafonte}, {Dapergolas}, {David}, {David}, {de Laverny}, {De Luise}, {De March}, {De Ridder}, {de Souza}, {de Torres}, {del Peloso}, {del Pozo}, {Delbo}, {Delgado}, {Delisle}, {Demouchy}, {Dharmawardena}, {Di Matteo}, {Diakite}, {Diener}, {Distefano}, {Dolding}, {Edvardsson}, {Enke}, {Fabre}, {Fabrizio}, {Faigler}, {Fedorets}, {Fernique}, {Fienga}, {Figueras}, {Fournier}, {Fouron}, {Fragkoudi}, {Gai}, {Garcia-Gutierrez}, {Garcia-Reinaldos}, {Garc{\'\i}a-Torres}, {Garofalo}, {Gavel}, {Gavras}, {Gerlach}, {Geyer}, {Giacobbe}, {Gilmore}, {Girona}, {Giuffrida}, {Gomel}, {Gomez}, {Gonz{\'a}lez-N{\'u}{\~n}ez},
  {Gonz{\'a}lez-Santamar{\'\i}a}, {Gonz{\'a}lez-Vidal}, {Granvik}, {Guillout}, {Guiraud}, {Guti{\'e}rrez-S{\'a}nchez}, {Guy}, {Hatzidimitriou}, {Hauser}, {Haywood}, {Helmer}, {Helmi}, {Sarmiento}, {Hidalgo}, {Hilger}, {H{\l}adczuk}, {Hobbs}, {Holland}, {Huckle}, {Jardine}, {Jasniewicz}, {Jean-Antoine Piccolo}, {Jim{\'e}nez-Arranz}, {Jorissen}, {Juaristi Campillo}, {Julbe}, {Karbevska}, {Kervella}, {Khanna}, {Kontizas}, {Kordopatis}, {Korn}, {K{\'o}sp{\'a}l}, {Kostrzewa-Rutkowska}, {Kruszy{\'n}ska}, {Kun}, {Laizeau}, {Lambert}, {Lanza}, {Lasne}, {Le Campion}, {Lebreton}, {Lebzelter}, {Leccia}, {Leclerc}, {Lecoeur-Taibi}, {Liao}, {Licata}, {Lindstr{\o}m}, {Lister}, {Livanou}, {Lobel}, {Lorca}, {Loup}, {Madrero Pardo}, {Magdaleno Romeo}, {Managau}, {Mann}, {Manteiga}, {Marchant}, {Marconi}, {Marcos}, {Marcos Santos}, {Mar{\'\i}n Pina}, {Marinoni}, {Marocco}, {Marshall}, {Martin Polo}, {Mart{\'\i}n-Fleitas}, {Marton}, {Mary}, {Masip}, {Massari}, {Mastrobuono-Battisti}, {Mazeh}, {McMillan}, {Messina}, {Michalik},
  {Millar}, {Mints}, {Molina}, {Molinaro}, {Moln{\'a}r}, {Monari}, {Mongui{\'o}}, {Montegriffo}, {Montero}, {Mor}, {Mora}, {Morbidelli}, {Morel}, {Morris}, {Muraveva}, {Murphy}, {Musella}, {Nagy}, {Noval}, {Oca{\~n}a}, {Ogden}, {Ordenovic}, {Osinde}, {Pagani}, {Pagano}, {Palaversa}, {Palicio}, {Pallas-Quintela}, {Panahi}, {Payne-Wardenaar}, {Pe{\~n}alosa Esteller}, {Penttil{\"a}}, {Pichon}, {Piersimoni}, {Pineau}, {Plachy}, {Plum}, {Poggio}, {Pr{\v{s}}a}, {Pulone}, {Racero}, {Ragaini}, {Rainer}, {Raiteri}, {Rambaux}, {Ramos}, {Ramos-Lerate}, {Re Fiorentin}, {Regibo}, {Richards}, {Rios Diaz}, {Ripepi}, {Riva}, {Rix}, {Rixon}, {Robichon}, {Robin}, {Robin}, {Roelens}, {Rogues}, {Rohrbasser}, {Romero-G{\'o}mez}, {Rowell}, {Royer}, {Ruz Mieres}, {Rybicki}, {Sadowski}, {S{\'a}ez N{\'u}{\~n}ez}, {Sagrist{\`a} Sell{\'e}s}, {Sahlmann}, {Salguero}, {Samaras}, {Sanchez Gimenez}, {Sanna}, {Santove{\~n}a}, {Sarasso}, {Schultheis}, {Sciacca}, {Segol}, {Segovia}, {S{\'e}gransan}, {Semeux}, {Shahaf}, {Siddiqui}, {Siebert},
  {Siltala}, {Silvelo}, {Slezak}, {Slezak}, {Smart}, {Snaith}, {Solano}, {Solitro}, {Souami}, {Souchay}, {Spagna}, {Spina}, {Spoto}, {Steele}, {Steidelm{\"u}ller}, {Stephenson}, {S{\"u}veges}, {Surdej}, {Szabados}, {Szegedi-Elek}, {Taris}, {Taylor}, {Teixeira}, {Tolomei}, {Tonello}, {Torra}, {Torra}, {Torralba Elipe}, {Trabucchi}, {Tsounis}, {Turon}, {Ulla}, {Unger}, {Vaillant}, {van Dillen}, {van Reeven}, {Vanel}, {Vecchiato}, {Viala}, {Vicente}, {Voutsinas}, {Weiler}, {Wevers}, {Wyrzykowski}, {Yoldas}, {Yvard}, {Zhao}, {Zorec}, {Zucker}, \& {Zwitter}}]{gaia_dr3}
{Gaia Collaboration}, {Vallenari}, A., {Brown}, A.~G.~A., {et~al.} 2023, \aap, 674, A1, \dodoi{10.1051/0004-6361/202243940}

\bibitem[{{Gaustad} {et~al.}(2001){Gaustad}, {McCullough}, {Rosing}, \& {Van Buren}}]{SHASSA}
{Gaustad}, J.~E., {McCullough}, P.~R., {Rosing}, W., \& {Van Buren}, D. 2001, \pasp, 113, 1326, \dodoi{10.1086/323969}

\bibitem[{{Haffner}(2010)}]{distribution_of_wim}
{Haffner}, L.~M. 2010, in Astronomical Society of the Pacific Conference Series, Vol. 438, The Dynamic Interstellar Medium: A Celebration of the Canadian Galactic Plane Survey, ed. R.~{Kothes}, T.~L. {Landecker}, \& A.~G. {Willis}, 179, \dodoi{10.48550/arXiv.1008.0622}

\bibitem[{{Haffner} {et~al.}(2003){Haffner}, {Reynolds}, {Tufte}, {Madsen}, {Jaehnig}, \& {Percival}}]{NORTHERN_WHAM}
{Haffner}, L.~M., {Reynolds}, R.~J., {Tufte}, S.~L., {et~al.} 2003, \apjs, 149, 405, \dodoi{10.1086/378850}

\bibitem[{{Haffner} {et~al.}(2009){Haffner}, {Dettmar}, {Beckman}, {Wood}, {Slavin}, {Giammanco}, {Madsen}, {Zurita}, \& {Reynolds}}]{warm_ionized_medium}
{Haffner}, L.~M., {Dettmar}, R.~J., {Beckman}, J.~E., {et~al.} 2009, Reviews of Modern Physics, 81, 969, \dodoi{10.1103/RevModPhys.81.969}

\bibitem[{{Haffner} {et~al.}(2010){Haffner}, {Reynolds}, {Madsen}, {Hill}, {Barger}, {Jaehnig}, {Mierkiewicz}, {Percival}, \& {Chopra}}]{SOUTHERN_WHAM}
{Haffner}, L.~M., {Reynolds}, R.~J., {Madsen}, G.~J., {et~al.} 2010, in Astronomical Society of the Pacific Conference Series, Vol. 438, The Dynamic Interstellar Medium: A Celebration of the Canadian Galactic Plane Survey, ed. R.~{Kothes}, T.~L. {Landecker}, \& A.~G. {Willis}, 388, \dodoi{10.48550/arXiv.1008.0612}

\bibitem[{Harris {et~al.}(2020)Harris, Millman, van~der Walt, Gommers, Virtanen, Cournapeau, Wieser, Taylor, Berg, Smith, Kern, Picus, Hoyer, van Kerkwijk, Brett, Haldane, del R{\'{i}}o, Wiebe, Peterson, G{\'{e}}rard-Marchant, Sheppard, Reddy, Weckesser, Abbasi, Gohlke, \& Oliphant}]{numpy}
Harris, C.~R., Millman, K.~J., van~der Walt, S.~J., {et~al.} 2020, Nature, 585, 357, \dodoi{10.1038/s41586-020-2649-2}

\bibitem[{{Hartmann}(1905)}]{Hartmann1905}
{Hartmann}, J. 1905, \apj, 21, 389, \dodoi{10.1086/141231}

\bibitem[{{H{\o}g} {et~al.}(2000){H{\o}g}, {Fabricius}, {Makarov}, {Urban}, {Corbin}, {Wycoff}, {Bastian}, {Schwekendiek}, \& {Wicenec}}]{tycho}
{H{\o}g}, E., {Fabricius}, C., {Makarov}, V.~V., {et~al.} 2000, \aap, 355, L27

\bibitem[{{Homa} {et~al.}(2023){Homa}, {Schiminovich}, {Putman}, {diCicco}, {Walker}, \& {Mittelman}}]{julia_aas_presentation}
{Homa}, J., {Schiminovich}, D., {Putman}, M., {et~al.} 2023, in American Astronomical Society Meeting Abstracts, Vol. 241, American Astronomical Society Meeting Abstracts, 223.04

\bibitem[{Hunter(2007)}]{matplotlib}
Hunter, J.~D. 2007, Computing in Science \& Engineering, 9, 90, \dodoi{10.1109/MCSE.2007.55}

\bibitem[{{Jacob} {et~al.}(2010){Jacob}, {Katz}, {Berriman}, {Good}, {Laity}, {Deelman}, {Kesselman}, {Singh}, {Su}, {Prince}, \& {Williams}}]{montagepy}
{Jacob}, J.~C., {Katz}, D.~S., {Berriman}, G.~B., {et~al.} 2010, {Montage: An Astronomical Image Mosaicking Toolkit}, Astrophysics Source Code Library, record ascl:1010.036

\bibitem[{{Lang} {et~al.}(2010){Lang}, {Hogg}, {Mierle}, {Blanton}, \& {Roweis}}]{astrometry_net}
{Lang}, D., {Hogg}, D.~W., {Mierle}, K., {Blanton}, M., \& {Roweis}, S. 2010, \aj, 139, 1782, \dodoi{10.1088/0004-6256/139/5/1782}

\bibitem[{Magnier {et~al.}(2020)Magnier, Schlafly, Finkbeiner, Tonry, Goldman, Röser, Schilbach, Casertano, Chambers, Flewelling, Huber, Price, Sweeney, Waters, Denneau, Draper, Hodapp, Jedicke, Kaiser, Kudritzki, Metcalfe, Stubbs, \& Wainscoat}]{panstarrs_calib}
Magnier, E.~A., Schlafly, E.~F., Finkbeiner, D.~P., {et~al.} 2020, The Astrophysical Journal Supplement Series, 251, 6, \dodoi{10.3847/1538-4365/abb82a}

\bibitem[{{McKee} \& {Ostriker}(1977)}]{ism_three_components}
{McKee}, C.~F., \& {Ostriker}, J.~P. 1977, \apj, 218, 148, \dodoi{10.1086/155667}

\bibitem[{{McKee} \& {Williams}(1997)}]{ob_associations}
{McKee}, C.~F., \& {Williams}, J.~P. 1997, \apj, 476, 144, \dodoi{10.1086/303587}

\bibitem[{{Mongui{\'o}} {et~al.}(2020){Mongui{\'o}}, {Greimel}, {Drew}, {Barentsen}, {Groot}, {Irwin}, {Casares}, {G{\"a}nsicke}, {Carter}, {Corral-Santana}, {Gentile-Fusillo}, {Greiss}, {van Haaften}, {Hollands}, {Jones}, {Kupfer}, {Manser}, {Murphy}, {McLeod}, {Oosting}, {Parker}, {Pyrzas}, {Rodr{\'\i}guez-Gil}, {van Roestel}, {Scaringi}, {Schellart}, {Toloza}, {Vaduvescu}, {van Spaandonk}, {Verbeek}, {Wright}, {Eisl{\"o}ffel}, {Fabregat}, {Harris}, {Morris}, {Phillipps}, {Raddi}, {Sabin}, {Unruh}, {Vink}, {Wesson}, {Cardwell}, {de Burgos}, {Cochrane}, {Doostmohammadi}, {Mocnik}, {Stoev}, {Su{\'a}rez-Andr{\'e}s}, {Tudor}, {Wilson}, \& {Zegmott}}]{IGAPS}
{Mongui{\'o}}, M., {Greimel}, R., {Drew}, J.~E., {et~al.} 2020, \aap, 638, A18, \dodoi{10.1051/0004-6361/201937333}

\bibitem[{{Morgan} {et~al.}(1955){Morgan}, {Str{\"o}mgren}, \& {Johnson}}]{Morgan1955}
{Morgan}, W.~W., {Str{\"o}mgren}, B., \& {Johnson}, H.~M. 1955, \apj, 121, 611, \dodoi{10.1086/146026}

\bibitem[{{Moss} {et~al.}(2012){Moss}, {McClure-Griffiths}, {Braun}, {Hill}, \& {Madsen}}]{supershell}
{Moss}, V.~A., {McClure-Griffiths}, N.~M., {Braun}, R., {Hill}, A.~S., \& {Madsen}, G.~J. 2012, \mnras, 421, 3159, \dodoi{10.1111/j.1365-2966.2012.20538.x}

\bibitem[{{Newton} {et~al.}(2017){Newton}, {Irwin}, {Charbonneau}, {Berlind}, {Calkins}, \& {Mink}}]{ha_emission_of_m_dwarf_rotation}
{Newton}, E.~R., {Irwin}, J., {Charbonneau}, D., {et~al.} 2017, \apj, 834, 85, \dodoi{10.3847/1538-4357/834/1/85}

\bibitem[{Núñez {et~al.}(2024)Núñez, Agüeros, Curtis, Covey, Douglas, Chu, DeLaurentiis, Wang, \& Drake}]{chromospheric_activity_rotation_praesepe_hyades}
Núñez, A., Agüeros, M.~A., Curtis, J.~L., {et~al.} 2024, The Astrophysical Journal, 962, 12, \dodoi{10.3847/1538-4357/ad117e}

\bibitem[{Ochsenbein(1996)}]{10.26093/cds/vizier}
Ochsenbein, F. 1996, The VizieR database of astronomical catalogues,  CDS, Centre de DonnÃ©es astronomiques de Strasbourg, \dodoi{10.26093/CDS/VIZIER}

\bibitem[{{Ochsenbein} {et~al.}(2000){Ochsenbein}, {Bauer}, \& {Marcout}}]{vizier}
{Ochsenbein}, F., {Bauer}, P., \& {Marcout}, J. 2000, \aaps, 143, 23, \dodoi{10.1051/aas:2000169}

\bibitem[{pandas~development team(2020)}]{pandas}
pandas~development team, T. 2020, pandas-dev/pandas: Pandas, latest,  Zenodo, \dodoi{10.5281/zenodo.3509134}

\bibitem[{{Pickering}(1891)}]{draper_catalog}
{Pickering}, E.~C. 1891, Annals of Harvard College Observatory, 26, 1

\bibitem[{{Raymond} {et~al.}(2020){Raymond}, {Caldwell}, {Fesen}, {Weil}, {Boumis}, {di Cicco}, {Mittelman}, \& {Walker}}]{galactic_halo_snr_sharp_ha_filaments}
{Raymond}, J.~C., {Caldwell}, N., {Fesen}, R.~A., {et~al.} 2020, \apj, 888, 90, \dodoi{10.3847/1538-4357/ab5e84}

\bibitem[{{Reynolds}(1984)}]{reynolds84}
{Reynolds}, R.~J. 1984, \apj, 282, 191, \dodoi{10.1086/162190}

\bibitem[{{Sabin} {et~al.}(2013){Sabin}, {Parker}, {Contreras}, {Olgu{\'\i}n}, {Frew}, {Stupar}, {V{\'a}zquez}, {Wright}, {Corradi}, \& {Morris}}]{igaps_sn_remnants}
{Sabin}, L., {Parker}, Q.~A., {Contreras}, M.~E., {et~al.} 2013, \mnras, 431, 279, \dodoi{10.1093/mnras/stt160}

\bibitem[{{Sharpless} \& {Osterbrock}(1952)}]{Sharpless1952}
{Sharpless}, S., \& {Osterbrock}, D. 1952, \apj, 115, 89, \dodoi{10.1086/145516}

\bibitem[{{Soderblom} {et~al.}(1993){Soderblom}, {Stauffer}, {Hudon}, \& {Jones}}]{chromospheric_emission_f_g_k_dwarfs}
{Soderblom}, D.~R., {Stauffer}, J.~R., {Hudon}, J.~D., \& {Jones}, B.~F. 1993, \apjs, 85, 315, \dodoi{10.1086/191767}

\bibitem[{{Stetson}(1987)}]{Stetson1987}
{Stetson}, P.~B. 1987, \pasp, 99, 191, \dodoi{10.1086/131977}

\bibitem[{{STScI}(2022)}]{PS_DR1_MAST_DOI}
{STScI}. 2022, Pan-STARRS1 DR1 Catalog,  STScI/MAST, \dodoi{10.17909/55E7-5X63}

\bibitem[{{Viironen} {et~al.}(2009){Viironen}, {Greimel}, {Corradi}, {Mampaso}, {Rodr{\'\i}guez}, {Sabin}, {Delgado-Inglada}, {Drew}, {Giammanco}, {Gonz{\'a}lez-Solares}, {Irwin}, {Miszalski}, {Parker}, {Rodr{\'\i}guez-Flores}, \& {Zijlstra}}]{candidate_pn_iphas_catalog}
{Viironen}, K., {Greimel}, R., {Corradi}, R.~L.~M., {et~al.} 2009, \aap, 504, 291, \dodoi{10.1051/0004-6361/200912002}

\bibitem[{Virtanen {et~al.}(2020)Virtanen, Gommers, Oliphant, Haberland, Reddy, Cournapeau, Burovski, Peterson, Weckesser, Bright, {van der Walt}, Brett, Wilson, Millman, Mayorov, Nelson, Jones, Kern, Larson, Carey, Polat, Feng, Moore, {VanderPlas}, Laxalde, Perktold, Cimrman, Henriksen, Quintero, Harris, Archibald, Ribeiro, Pedregosa, {van Mulbregt}, \& {SciPy 1.0 Contributors}}]{scipy}
Virtanen, P., Gommers, R., Oliphant, T.~E., {et~al.} 2020, Nature Methods, 17, 261, \dodoi{10.1038/s41592-019-0686-2}

\bibitem[{{Witham} {et~al.}(2008){Witham}, {Knigge}, {Drew}, {Greimel}, {Steeghs}, {G{\"a}nsicke}, {Groot}, \& {Mampaso}}]{iphas_cat_ha_emission_sources}
{Witham}, A.~R., {Knigge}, C., {Drew}, J.~E., {et~al.} 2008, \mnras, 384, 1277, \dodoi{10.1111/j.1365-2966.2007.12774.x}

\end{thebibliography}


\end{document}